\documentclass[useAMS,usenatbib]{mn2e}

\usepackage{epsfig,amsmath,amssymb}
\usepackage{natbib}
\usepackage{rotating}
\usepackage{longtable,lscape}

\def \mgii {Mg\,{\textsc {ii}}}
\def \aliii {Al\,{\textsc {iii}}}
\def \civ {C\,{\textsc {iv}}}
\def \siiv {Si\,{\textsc {iv}}}

\def \oiii {[O\,{\textsc {iii}}]}

\def \nv {N\,{\textsc {v}}}

\def \ebv {\ensuremath{E(B-V)}}

\def \md {\ensuremath{\bar{d}}}

\def \kms{\ensuremath{{\rm km\,s}^{-1}}}

\let\oldsqrt\sqrt
\def\sqrt{\mathpalette\DHLhksqrt}
\def\DHLhksqrt#1#2{
\setbox0=\hbox{$#1\oldsqrt{#2\,}$}\dimen0=\ht0
\advance\dimen0-0.3\ht0
\setbox2=\hbox{\vrule height\ht0 depth -\dimen0}
{\box0\lower0.4pt\box2}}

\begin{document}

\title[A strong redshift dependence of the BALQSO fraction]{A strong
  redshift dependence of the broad absorption line quasar fraction}

\author[J.~T.~Allen et al.]{James T. Allen,$^1$\thanks{E-mail:
    jta@ast.cam.ac.uk} Paul C. Hewett$^1$, Natasha Maddox$^2$, Gordon
  T. Richards$^3$ \newauthor and Vasily Belokurov$^1$\\
  $^1$Institute of Astronomy, University of Cambridge, Madingley Road,
  Cambridge CB3 0HA \\
  $^2$Astrophysikalisches Institut Potsdam, An der Sternwarte 16,
  14482 Potsdam, Germany \\
  $^3$Department of Physics, Drexel University, Philadelphia, PA
  19104, USA}

\maketitle

\begin{abstract}
We describe the application of non-negative matrix factorisation to
generate compact reconstructions of quasar spectra from the Sloan
Digital Sky Survey (SDSS), with particular reference to broad
absorption line quasars (BALQSOs).  BAL properties are measured for
\siiv\ $\lambda 1400$, \civ\ $\lambda 1550$, \aliii\ $\lambda 1860$
and \mgii\ $\lambda 2800$, resulting in a catalogue of 3547 BALQSOs.
Two corrections, based on extensive testing of synthetic BALQSO
spectra, are applied in order to estimate the intrinsic fraction of
\civ\ BALQSOs.  First, the probability of an observed BALQSO spectrum
being identified as such by our algorithm is calculated as a function
of redshift, signal-to-noise ratio and BAL properties.  Second, the
different completenesses of the SDSS target selection algorithm for
BALQSOs and non-BAL quasars are quantified.  Accounting for these
selection effects the intrinsic \civ\ BALQSO fraction is 41$\pm$5 per
cent.  Our analysis of the selection effects allows us to measure the
dependence of the intrinsic \civ\ BALQSO fraction on luminosity and
redshift.  We find a factor of 3.5$\pm$0.4 decrease in the intrinsic
fraction from the highest redshifts, $z$$\simeq$4.0, down to
$z$$\simeq$2.0.  The redshift dependence implies that an orientation
effect alone is not sufficient to explain the presence of BAL troughs
in some but not all quasar spectra.  Our results are consistent with
the intrinsic BALQSO fraction having no strong luminosity dependence,
although with 3-$\sigma$ limits on the rate of change of the intrinsic
fraction with luminosity of $-$6.9 and 7.0 per cent dex$^{-1}$ we are
unable to rule out such a dependence.
\end{abstract}


\section{Introduction}

Despite the greatly increased quantity and quality of data since the
first observations of broad absorption line quasars (BALQSOs; see
\citealt*{Weymann81} for a review), relatively little is known about
their physical origins.  By definition, a BALQSO exhibits blueshifted
absorption, which may extend to tens of thousands of kilometres per
second relative to the quasar systemic velocity, with an absorption
velocity width of at least 2000\,\kms\ \citep{Weymann81}.  Such
broad absorption features with comparably large redshifted velocities
are not observed, implying that the absorption originates in material
outflowing from the quasar.

It is not yet clear whether the material outflowing in BAL systems
represents a large mass relative to outflow processes as a whole, with
estimates of the total column density of absorbing material varying by
several orders of magnitude (\citealt{Hamann98}; \citealt*{Gabel06};
\citealt{Casebeer08,Hamann08}).  However, the absorption can certainly
be used to trace the kinematics of the quasar outflows.  As outflows
are believed to be fundamental to the phenomenon of active galactic
nuclei (AGN) feedback a better understanding of their kinematics could
shed light on the physical basis of the outflow phenomenon (see
\citealt{Begelman04} for a review of AGN feedback mechanisms).  In
some models the BAL phenomenon arises from preferred viewing angles
towards the central regions and the demographics of BALQSOs among the
quasar population may also help constrain the parameters of unified
models (e.g.\ \citealt{Elvis00}).

Broad absorption is most commonly observed in high-ionisation lines
such as \civ\ $\lambda$1550 and \siiv\ $\lambda$1400; BALQSOs
without absorption present in lower-ionisation species are termed
HiBALQSOs.  Less frequently, broad absorption is also observed in the
low-ionisation lines \aliii\ $\lambda$1860 and \mgii\ $\lambda$2800.
BALQSOs with both high- and low-ionisation absorption are termed
LoBALQSOs, while those even rarer objects that also show absorption by
one or more species of iron are termed FeLoBALQSOs. See \citet{Hall02}
for examples of the rich variety of spectra produced by the
broad absorption phenomenon.

The most widely used metric to separate BALQSOs and non-BAL quasars is the
balnicity index (BI), first presented by \citet{Weymann91}.  The BI is
defined as
\begin{equation}
\label{eq:bi}
{\rm BI} = \int_{-25000\,\kms}^{-3000\,\kms}
\left( 1 - \frac{f(v)}{0.9} \right) C {\rm d}v,
\end{equation}
where $f(v)$ is the continuum-normalised flux as a function of
velocity, $v$, relative to the line centre.  The constant $C$ is equal
to unity in regions where $f(v)$ has been continuously less than 0.9
for at least 2000\,\kms, counting from large to small outflow
velocities, and zero elsewhere.  The outflow velocity, $v$, is defined
to be negative for blueshift velocities, i.e. for material moving
towards the observer.  Quasars with non-zero BI are defined as
BALQSOs; all others are non-BAL quasars.

Modifications to the BI have been proposed, in particular the
absorption index (AI; \citealt{Hall02,Trump06}), designed to include
narrower troughs than the BI, and the modified balnicity index BI$_0$
(\citealt{Gibson09}, hereafter G09), which extends the integration
region to zero velocity.  Each of these modifications
increases the number of quasars defined to be BALQSOs, but
\citet{Knigge08} demonstrated that the observed distribution of AI
values is bimodal, suggesting that the metric encompasses two
distinct populations of absorbers.  Except where noted we define
BALQSOs in this paper according to the traditional BI.

The {\it observed} \civ\ BALQSO fraction has been variously calculated
as, among other measurements, 15$\pm$3 \citep{HF03}, 14.0$\pm$1.0
\citep{Reichard03b}, 12.5 \citep{Scaringi09}, and 13.3$\pm$0.6 per
cent (G09).  Correcting for the different probabilities of a BALQSO
and non-BAL quasars entering the spectroscopic surveys used, the
intrinsic fraction present in flux-limited optical surveys has been
estimated as 22$\pm$4 \citep{HF03}, 15.9$\pm$1.4 \citep{Reichard03b},
17$\pm$3 \citep{Knigge08}, and 16.4$\pm$0.6 per cent (G09).

It is often suggested that broad absorption occurs in all quasars but
is only observed along particular sightlines
(e.g. \citealt{Weymann91,Elvis00}).  In this model the BALQSO fraction
can be directly interpreted as the solid angle covered by the
absorbing clouds divided by the solid angle over which a Type 1 quasar
can be seen.  Another possibility is that BALQSOs could represent a
particular evolutionary stage (e.g.\ \citealt*{Voit93};
\citealt{Becker97,Lipari06}), during which absorbing material with a
high covering fraction is being expelled from the central regions of
the quasar.  However, models in which individual BALQSOs have a very
high covering fraction appear to be ruled out by the results of
\citet{Gallagher07}, which show that BAL and non-BAL quasars have very
similar mid-infrared properties, while a high covering fraction would
result in more light being reprocessed into the mid-infrared.  The
possibility still remains that a combination of evolutionary and
orientation effects can explain the separation of BALQSOs and non-BAL
quasars.  For example, in a disc wind model the structure of the wind
could change with cosmic time, while still having an orientation
dependence.

It is difficult or impossible to distinguish between the different
models based on observations of individual objects, but much progress
can be made by characterising the statistical properties of the BALQSO
population.  Crucial properties in such an investigation are the
dependence of the BALQSO fraction on factors such as redshift and
luminosity.  However, measuring these properties is greatly
complicated by the selection effects involved, which are themselves
strongly dependent on redshift and luminosity, in part due to their
correlations with the signal-to-noise ratio (S/N) of observed
spectra.  For example, the probability of correctly identifying a
BALQSO increases with increasing S/N, which is in turn related to the
luminosity of an observed quasar.  Without quantifying the form of the
selection effect it is impossible to establish whether or not an {\em
  observed} trend in the BALQSO fraction with luminosity is the result
of an {\em intrinsic} trend.  The existence of such an effect has
previously been noted (\citealt{Knigge08}; G09) but not fully
addressed; we quantify the S/N-dependent selection probability and
other selection effects in this paper.

A pre-requisite for any investigation of the statistical distribution
of absorption properties is a large well-defined catalogue of BALQSOs.
The Sloan Digital Sky Survey (SDSS; \citealt{SDSS}) is very well suited
to this purpose as it provides a large homogeneous sample of quasar
spectra in which to search for BALQSOs.  Previous studies producing
BALQSO catalogues from the SDSS include \citet{Reichard03a}, using the
Early Data Release (EDR), \citet{Trump06}, using the Third Data
Release (DR3), and G09 and \citet{Scaringi09}, both using DR5. In this
work, quasars from the SDSS DR6 are used, giving a larger sample size.

The BI, AI and BI$_0$ are all calculated from the continuum-normalised
flux, so all require an estimate of the unabsorbed continuum level.
Improving the quality of the continua will greatly improve the
accuracy of the resulting determinations of statistical properties, as
the continuum level is currently the principal uncertainty in the
classification of BALQSOs.  Previous studies have estimated the
continuum using template spectra \citep{Trump06} or simple continuum
plus emission-line models (G09), but each of these techniques has its
disadvantages: template spectra are limited in the range of continua
and emission-line profiles they allow, while any technique that relies
on directly fitting the emission lines will struggle when significant
sections of those emission lines are absorbed.

In this paper we produce estimates of the unabsorbed emission using
non-negative matrix factorisation (NMF; \citealt{LS99, LS00, BR07}), a
blind source separation technique.  NMF uses all the available
information to fit the entire spectrum simultaneously, enabling
accurate reconstructions to be produced even in cases where much of
the spectrum has been absorbed.  Starting from a suitably chosen input
sample the technique is able to reconstruct spectra with a wide range
of observed properties with little or no manual intervention.  The
estimates of the continua in the \civ\ BI region will be made publicly
available through the SDSS value-added catalogues as this paper is
published, to enable their use in any future BALQSO classification
schemes.

In Section~\ref{sc:bss} we introduce blind source separation and
non-negative matrix factorisation.  In Section~\ref{sc:sample} we
present our sample of SDSS quasars, and in
Section~\ref{sc:reconstructions} we describe our method of
reconstructing their spectra.  The results are presented in
Section~\ref{sc:results}, including the resulting catalogue of
BALQSOs.  In Section~\ref{sc:comparison} we compare these results to
those from previous studies.  In Sections \ref{sc:p_det} and
\ref{sc:selection} we quantify the probability of a BAL trough being
detected by our methods, and the relative probabilities of BALQSOs and
non-BAL quasars entering the SDSS spectroscopic survey.  From these
results the intrinsic BALQSO fraction is calculated in
Section~\ref{sc:intrinsic}.  The results are discussed in
Section~\ref{sc:discussion}, and we summarise our conclusions in
Section~\ref{sc:conclusions}.

Those readers who are particularly interested in the BALQSO catalogue
itself, its generation and the observed properties of BALQSOs should
focus on Sections \ref{sc:sample}--\ref{sc:results}.  Readers who are
more interested in the relation between the observed and intrinsic
properties of the BALQSO population, and in particular the calculation
of the intrinsic fraction of BALQSOs as a function of redshift and
luminosity, can skip to Section~\ref{sc:p_det}.

In this work we assume a flat $\Lambda$CDM cosmology with
$H_0$=70\,\kms\,Mpc$^{-1}$, $\Omega_{\rm M}$=0.3 and
$\Omega_\Lambda$=0.7.  Vacuum wavelengths are employed throughout the
paper.

\section{Blind source separation}

\label{sc:bss}

Blind source separation (BSS) techniques involve rewriting an $n
\times m$ data matrix $V$ as the product of a set of components, $H$,
and weights, $W$:
\begin{equation}
\label{eq:vwh}
V = W H.
\end{equation}
In the context of this work, $V$ is an $n \times m$ array of flux
measurements for $n$ different quasars at $m$ wavelengths, $H$ is an
$r \times m$ array of the $r$ component spectra at the same
wavelengths, and $W$ is an $n \times r$ array of the corresponding
weights for each quasar.  For any individual quasar, the
observed spectrum is written as a linear combination of the $r$
components.  In the case that $r < n$, the equality in
equation~\ref{eq:vwh} is an approximation, and the product $WH$ can be
viewed as a reconstruction of the original data.  It is this
reconstruction that we use as an estimate of the unabsorbed continuum
of BALQSOs.

\subsection{Non-negative matrix factorisation}

\label{ss:nmf}

Non-negative matrix factorisation (NMF) is a relatively new BSS
technique that incorporates a non-negativity constraint on both its
components and their weights \citep{LS99,LS00,BR07}.  The
non-negativity constraint is appealing in the context of
spectroscopic data as the physical emission signatures are expected to
obey such a restriction naturally.  Unusually for a BSS technique,
fewer components are generated than there are input spectra.  Starting
from random initial matrices, the components, $H$, and weights, $W$,
follow the multiplicative update rules:
\begin{equation}
\label{eq:wupdate}
W_{ik} \leftarrow W_{ik} \frac{[VH^T]_{ik}}{[WHH^T]_{ik}},
\end{equation}
and
\begin{equation}
\label{eq:hupdate}
H_{kj} \leftarrow H_{kj} \frac{[W^TV]_{kj}}{[W^TWH]_{kj}},
\end{equation}
in which, at each step, the elements of $W$ and $H$ are replaced by the
values given by the right hand sides of equations
\ref{eq:wupdate} and \ref{eq:hupdate}.  The matrices continue to be
updated until a stable solution is reached.

The update rules minimize the error in the reconstructions, $WH$, as
measured by the Euclidean distance to the data, defined as
\begin{equation}
\label{eq:euc_dist}
\parallel V - W H \parallel = \sqrt{\sum_{ij}\left(V_{ij} - \sum_{k}
  W_{ik} H_{kj}\right)^2}.
\end{equation}
The random starting conditions result
in slightly different components being generated each time the
algorithm is executed.  However, as the Euclidean distance is defined
in terms of the reconstructions, and the algorithm converges on a
minimum in the Euclidean distance \citep{LS99,LS00}, the resulting
reconstructions do not vary.

After an initial source separation has been performed, generating a
set of components, the components can be applied to create compact
reconstructions of further spectra.  A new data matrix is created from
the spectra that are to be reconstructed, the components matrix
is taken from the initial results, and a random set of starting weights 
is generated.  The matrix of weights is then updated according to
equation~\ref{eq:wupdate}, keeping the components matrix fixed.  This
process finds a local minimum in the Euclidean distance between the
data and the reconstructions, under the constraint of fixed $H$.

\section{Quasar sample}\label{sc:sample}

The quasar sample consists of 91\,665 objects from the SDSS DR6
spectroscopic survey, including 77\,392
quasars in the \citet{Schneider07} DR5 quasar catalogue that
are retained in the later DR7 quasar catalogue of
\citet{Schneider10}.  A further 13\,081 objects are quasars,
present in the additional DR6 spectroscopic plates, identified by one
of us (PCH) using a similar prescription to that employed by
\citet{Schneider07}, all of which are present in the
\citet{Schneider10} catalogue.  An additional 1192 objects,
which do not satisfy one, or both, of the emission line velocity width
or absolute magnitude criterion imposed by
\citet{Schneider07}, are also included.  While formally
failing the `quasar' definition of Schneider et al.'s DR5 and DR7
compilations the objects are essentially all luminous active galactic
nuclei (AGN). In fact, the vast majority of such objects lie at redshifts
$z$$<$0.4, where even \mgii\ broad absorption is not visible in the SDSS 
spectra.  In the following work, objects with $z$$<$0.4 were not
searched for BAL features, and a further 30 spectra were discarded
because they were very poor quality or unavailable in DR6, leaving
93\,400 spectra of 86\,773 unique objects.

The spectra were all processed through the sky-residual subtraction scheme of
\citet{WH05b}, resulting in significantly improved S/N at observed
wavelengths $\lambda$$>$7200\,\AA.

The SDSS DR6 contains a very large number of objects for which multiple spectra
are available.  For our quasar sample there are $\simeq$9000 independent pairs
of spectra.  The catalogue of spectrum pairs allows an empirical
check on the reproducibility of the BI determinations as a function of 
spectrum S/N.

Improved quasar redshifts were taken from a preliminary implementation
of the \citet{HW10} scheme for generating
self-consistent redshifts for quasars with SDSS spectra. While not
responsible for major differences in the resulting catalogue of
BALQSOs, the large number of redshift revisions at the $\sim$1000\,\kms\
level does result in classification changes for a number of quasars
with absorption present at low velocities, $<$5000\,\kms, relative to
the quasar systemic redshift.

The SDSS spectra as provided are not corrected for the effects of
Galactic extinction, so before use we applied a correction using
the SDSS \ebv\ measurement, taken from the dust maps of
\citet*{SFD98}, and the Milky Way extinction curve of \citet*{CCM89}.

\section{Continuum reconstructions of quasars}

\label{sc:reconstructions}

In order to measure the BI, or any related metric, an estimate of the
unabsorbed emission from the quasar is required. Here, non-negative
matrix factorisation (NMF) is used to create sets of component spectra
from non-BAL quasars. These components are then used to reconstruct
the emission for all quasars, thereby allowing the implementation of
a BI-determination algorithm to define the sample of BALQSOs. The
implementation of the method is described in detail in the following
sections.

For any set of spectra, NMF is best applied to the common rest-frame
wavelength range.  To maximise the possible wavelength range, and to
allow for redshift evolution of quasar properties, the quasars in the
sample defined in Section~\ref{sc:sample} were divided into
non-overlapping redshift bins of width $\Delta z$=0.1.  Except where
noted, the methods described in this section were applied separately
to each such redshift bin.

\subsection{Quasar spectrum selection for component generation}

\label{ss:input}

To generate component spectra a sample of 500 input quasars were
selected in each redshift bin.  The input spectra were chosen to
possess spectrum S/Ns, specifically, ({\tt
SN\_R}+{\tt SN\_I})/2, within the restricted range 9--25 and to
possess at least 3800 `good' SDSS pixels. Quasars exhibiting the
presence of any form of extended absorption in their spectra longward
of Ly$\alpha$ $\lambda$1216 were then excluded. A simple, very
conservative, algorithm, based on the `bending' of absorption-free
template spectra to `fit' each spectrum, allowed the calculation of a
generic `absorption equivalent width'.  The S/N and absorption cuts
removed virtually all heavily-reddened quasars, but some additional
objects in the redshift bin 0.7$\le$$z$$<$0.8, with strong
\oiii\ $\lambda$5007 emission and heavily dust-reddened continua, were
manually removed.  The form of the NMF components
generated from the input quasar sample did not depend on the exact
definition of the sample of input spectra.

The spectra at high redshifts were truncated at 1175\,\AA, below which
wavelength the spectra are dominated by the Ly$\alpha$ forest,
severely attenuating the underlying quasar emission.  At redshifts
$z$$>$2.1 fewer than 500 quasars in each redshift bin satisfied the
selection requirements, so smaller input samples were used, reaching a
minimum of 202 quasars for 2.5$\le$$z$$<$2.6.  For $z$$\ge$2.6 there
were insufficient quasars to produce representative components. However, 
due to the imposition of the truncation at 1175\,\AA, the components 
generated from the 2.5$\le$$z$$<$2.6 interval covered the full accessible 
wavelength range present in quasars with $z$$>$2.6. Thus, the same set
of NMF components were used for all quasars with $z$$>$2.5.

In each redshift bin, NMF components were generated from the input
quasar spectra after pixels affected by any narrow absorption systems
present had been flagged using a median filter-based algorithm. Flux
values for such pixels, along with those pixels flagged as `bad' due
to a value of zero in the SDSS spectrum noise array, were generated
via interpolation. Similarly, 12.0-\AA\ regions centred on the
prominent sky lines at observed-frame wavelengths 5578.5 and
6301.7\,\AA\ were interpolated over.  Note that the use of spectra
with a very high fraction of `good' pixels ensured that interpolation
was applied for, at most, only a few, narrow, intervals in each
spectrum. The spectra were then normalised to have the same median
flux.

\subsection{Number of components}

In applying NMF the number of components to generate must be chosen
manually.  In all cases using more components will reduce the total
Euclidean distance between the reconstructions and the data, but above
a certain number the improvement comes from fitting to the noise of
individual spectra rather than to the general emission properties of
the quasars, giving no further improvement in the quality of
subsequent reconstructions.  An estimate of the optimal number of
components to employ can be found via the $\chi_\nu^2$ values of the
reconstructions of the input spectra: when too many components are
used and the NMF procedure is repoducing the noise from just one or a
few spectra, the `overfitted' spectra have much lower $\chi_\nu^2$ values
than any others in the input sample.

The point at which overfitting occurs is a function of both the S/N
and the emission properties of the input spectra.  The number of
components to use was chosen as the greatest number for which no input
spectra showed the drop in $\chi_\nu^2$ typical of overfitting.  In
practice the correct number of components could be selected from a
visual inspection of the $\chi_\nu^2$ distributions.  The
number of components so identified varied between 8 and 14 depending
on the redshift bin.

In order to reconstruct a small fraction of the quasar spectra the
number of components was reduced below the number initially
determined.  These cases are described in Sections \ref{ss:dipsearch}
and \ref{ss:manmask}.

\subsection{Automated masking of spectra}

The sets of components, derived as described above, were then applied
to generate reconstructions of all quasars in the sample, following
the procedure described in Section~\ref{ss:nmf}.  As reconstructions
of the unabsorbed emission were required, all wavelength regions
where absorption was present were masked during the fits, along with
regions affected by prominent sky lines. The same
recipe for identifying narrow absorption and sky lines described in
Section \ref{ss:input} was 
applied, but the exclusion of wavelength regions affected by broad 
absorption required an iteratively updated mask, as described below.

\begin{figure}
\includegraphics[width=84mm]{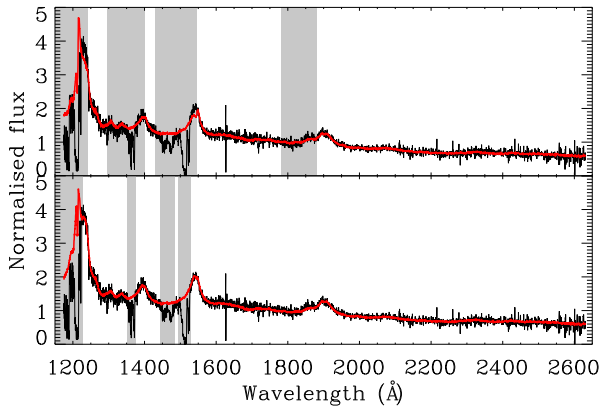}
 \caption{Illustration of the automatic masking procedure.  The upper
   panel shows the initial masked regions (grey shading) and the
   resulting NMF reconstruction (red line) of the observed flux
   (black line).  The lower panel shows the updated mask derived
   from a comparison of the initial reconstruction and the data.
   The new reconstruction, derived using the updated mask, is also 
   shown.  The object illustrated is SDSS J080559.94+140530.4.}
 \label{fg:automask}
\end{figure}

As an initial mask the wavelength regions $\lambda$$\le$1240,
1295$\le$$\lambda$$\le$1400, 1430$\le$$\lambda$$\le$1546, and
1780$\le$$\lambda$$\le$1880 (all in \AA) were removed, and a
reconstruction was generated based on the pixels remaining.  For
subsequent iterations the mask was re-defined by examining the data
and reconstructions within a window of width 31 pixels
centred on each pixel in turn: a pixel was masked if the majority of
pixels in its window had an observed flux lower than the
reconstruction by at least twice the noise level.  Any wavelength
regions masked in this way were extended by a radius of 10 pixels
to ensure the wings of the absorption were fully covered.
Information about the locations of previous masks was not used in
generating new masks and no restrictions were placed on potential mask
locations.  In most cases the mask locations converged in only a few
iterations.  The masking process is illustrated in
Fig.~\ref{fg:automask}, which shows the initial mask and the mask
after the first iteration for an example BALQSO.

\subsection{Accounting for quasar SED changes due to dust absorption}

\label{ss:slopecor}

NMF, like most BSS techniques, assumes the observed data is a linear
sum of various components.  In contrast, the extinction and reddening
of light by intervening dust, either within the quasar's host galaxy
or in an intervening absorber, has the effect of multiplying the
observed spectrum by a wavelength-dependent factor.

\begin{figure}
\includegraphics[width=84mm]{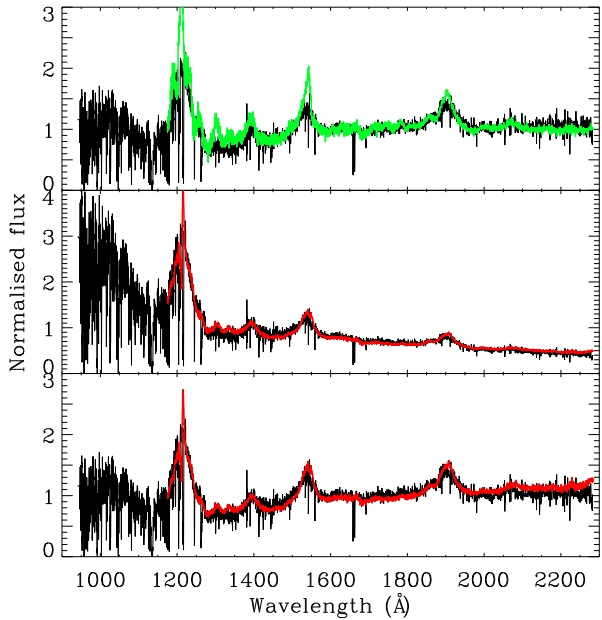}
 \caption{Top panel: observed flux (black) and initial NMF
   reconstruction (green) of SDSS J145907.19+002401.2, a quasar at
   $z$=3.04.  The quality of the fit is poor because the quasar is
   heavily reddened.  Middle panel: flux (black) of the same quasar
   after being divided by a power law slope with index 2.13, and the
   resulting NMF reconstruction (red).  Bottom panel: as middle panel,
   but after the flux and reconstruction have been multiplied by the
   same power law slope to restore the true shape of the observed quasar
   spectrum.}
 \label{fg:slopecor}
\end{figure}

For moderate levels of dust this effect is accounted for within the
NMF components, as these were generated from quasar spectra that were
themselves subject to varying levels of dust absorption. However, for
more strongly reddened objects, exhibiting dramatically steeper
spectral energy distributions (SEDs), an empirical correction was
applied to estimate the unreddened spectrum.  The method is
illustrated in Fig.~\ref{fg:slopecor}.  A composite quasar spectrum
was generated in each redshift bin from all quasars that satisfied the
input criteria listed in Section~\ref{ss:input}.  The ratio of each
observed spectrum to its corresponding composite spectrum was taken,
and a power law was fitted to the ratio to determine the level of
reddening.  Prominent emission lines were masked out during the fit.
If the index of the best-fitting power law was greater than 1.5,
indicating the observed spectrum was significantly redder than the
composite, the observed spectrum was divided by the best-fitting power law
to give the observed spectrum a similar shape to the composite.  An
NMF fit was then generated for the empirically `de-reddened' quasar
spectrum, and the fit was then multiplied by the best-fitting power law to
match the original observed spectrum\footnote{For redshifts $z$$<$0.6
the de-reddening procedure was not used as the observed quasar spectra
show a significantly greater spread in power law slopes, prohibiting a
clean separation of red outliers.}.

It should be emphasized that the procedure is in no way designed to 
parametrize, or otherwise quantify, the effect of dust on the quasars.
Rather, the procedure has been crafted simply to allow effective 
NMF reconstructions of quasars that possess very different SED shapes
compared to the bulk of the population.  The success of the power-law
normalisation can be attributed to the typical reddening towards SDSS
quasars being very close to a power law in form (\citealt{Hopkins04};
Maddox et~al.\ 2010, in preparation).

\subsection{Maximising C\,{\sevensize\bf IV} BAL trough coverage}

In taking the common wavelength range of a set of spectra spread over
a non-zero redshift range, some information from the ends of each observed
spectrum is necessarily discarded.  For quasars with $z$$\sim$1.6 the
\civ\ BAL trough at the blue end of the observed SDSS spectrum can be lost
as a result. To extend the ability to identify BALQSOs to the lowest
possible redshifts the NMF reconstructions of all quasars with
1.5$\le$$z$$<$1.7 were derived using the components from the next
highest redshift bins, i.e. quasars with 1.5$\le$$z$$<$1.6 used the
components from 1.6$\le$$z$$<$1.7, and quasars with 1.6$\le$$z$$<$1.7
used the components from 1.7$\le$$z$$<$1.8.

Using higher-redshift
components ensured maximal coverage of the \civ\ regions at the cost
of discarding a greater portion of the red end of the observed
spectra.  A similar scheme using lower-redshift components to cover a
BAL trough at the red end of a spectrum was not required as the
highest-redshift set of components used (2.5$\le$$z$$<$2.6) had
complete coverage of the \civ\ BAL region.  

\begin{figure}
\includegraphics[width=84mm]{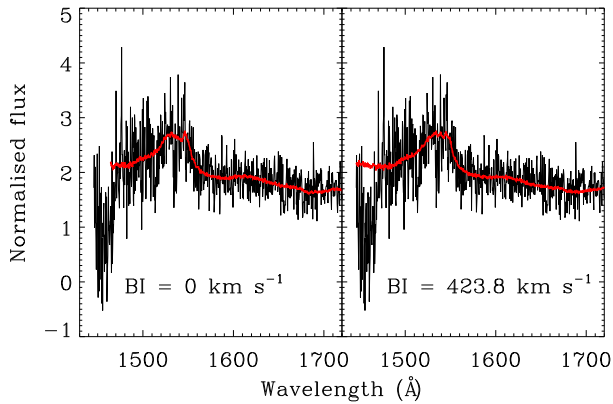}
 \caption{SDSS J143735.50+222340.3, a BALQSO at $z=1.64$.  In each
   panel the thin black line is the observed flux and the thick red
   line is the NMF reconstruction.  The NMF components for
   1.6$\le$$z$$<$1.7 do not extend below a wavelength of 1464\,\AA, so
   the reconstruction does not cover the BAL trough (left panel).
   Using the NMF components derived from 1.7$\le$$z$$<$1.8 quasars
   (right panel) allows the BAL trough to be identified.}
 \label{fg:zshift}
\end{figure}

An example BALQSO spectrum
with $z$=1.64 is shown in
Fig.~\ref{fg:zshift}, which shows how the BAL trough is only
identified when the higher-redshift components are used. The ability to 
detect BAL troughs at the extreme edges of spectra is a particular
advantage of the NMF-based technique.

\subsection{Rejecting unphysical emission line profiles}

\label{ss:dipsearch}

\begin{figure}
  \includegraphics[width=84mm]{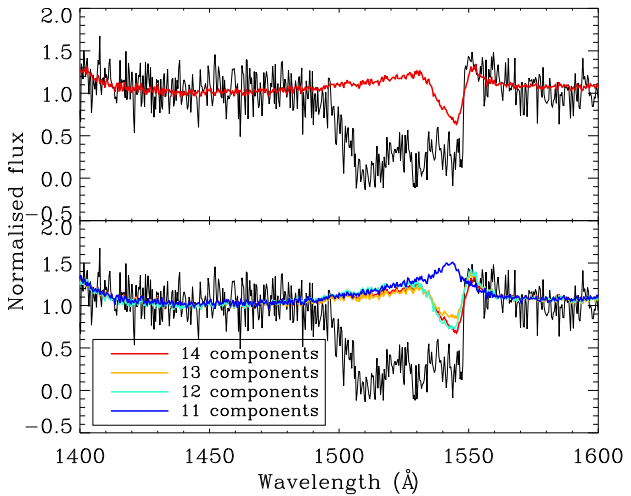}
  \caption{Top panel: Flux spectrum (black) of SDSS
    J104945.36+285823.3 with initial 14-component NMF
    reconstruction (red).  The initial reconstruction was identified
    as having an unphysical emission line profile.  Bottom panel: the
    same quasar spectrum (black) and the NMF reconstructions after
    additional masking, for 14, 13, 12 and 11 NMF components.  At 11
    components a satisfactory emission line profile is found and this
    reconstruction was adopted.}
  \label{fg:dipsearch}
\end{figure}

Although the automated technique described above produced excellent
reconstructions of both the continuum and emission lines for most
spectra, in up to 5 per cent of spectra an unphysical `dip' could be
identified in the profile of the \civ\ emission line
(Fig.~\ref{fg:dipsearch}).  Spectra where such artifacts were present
in the NMF reconstructions were identified via a simple `minima
detection' algorithm. Much as for the scheme used to generate an
absorption equivalent width (Section \ref{ss:input}), a template
spectrum was `bent' to fit the overall shape of the NMF reconstruction
and a simple scheme was used to identify any location where the NMF
reconstruction fell below the template spectrum.

In many cases the dip in the reconstructed emission line was caused by
incomplete masking of absorption in the observed spectrum, so spectra
with an identified dip were reprocessed with an additional mask
applied to the affected region.  The resulting fits were examined by
eye and accepted if the dip was no longer visible.  Where a dip was
still apparent in the reconstruction the number of components used to
create the reconstructions was reduced one by one until the dip was
eliminated.  For this purpose smaller sets of NMF components were
generated using the same input spectra as used to generate the initial 
base NMF sets.

Reducing the number of components reduces the freedom of
the fitting procedure to create unphysical line profiles, eliminating
the dips in almost all cases.  Quasars whose reconstructions
still exhibited a dip in the \civ\ emission line with as few as two
components are discussed further in the following section.

\subsection{Manual masking of spectra}

\label{ss:manmask}

Spectra with reconstructions that met one or more of the following
criteria were examined by eye to determine if the automated masking
had failed to correctly separate absorbed and unabsorbed regions:
\begin{enumerate}
  \item dip in \civ\ emission line profile still present with only two
    NMF components
  \item $\chi_\nu^2 \ge\ 2$
  \item at least 500 pixels masked
\end{enumerate}
Together these criteria selected $\simeq$2.5 per cent of the
reconstructions, of which $\simeq$20 per cent ($\simeq$0.5 per cent of
the total) were judged to require manually-defined
masks\footnote{Additionally, the automated procedure had failed to produce
fits for three spectra: for one the automated mask covered the entire
spectrum; for the second, the power law slope correction, as described
in Section~\ref{ss:slopecor}, was so great it suppressed all
information in the spectrum; for the third, large sections of the SDSS
spectrum had recorded fluxes of $\pm \infty$.  Manually-defined masks
were also created for these three spectra.}.

New reconstructions were generated for the selected spectra with
manually-defined masks used in place of the automated masks.  The
results were visually inspected to identify unphysical emission line
profiles and, where these were found, the number of components was
reduced as in Section~\ref{ss:dipsearch}.  At this point
reconstructions with only two components were allowed.  In a very
small number of cases the reconstructed emission line profiles still
exhibited unphysical profiles with as few as two components; in these
cases the best reconstruction available was chosen even if a dip was
present.

Notwithstanding the presence of the occasional pathological case where
the automated NMF continuum generation fails, the success of the NMF-based
scheme is evident from the extremely low percentage of quasars (0.5 per cent)
that require any manual intervention and that only 54 spectra posses
reconstructions that we judge to be significantly sub-optimal.

\subsection{Visual inspection of spectra}

As a final check, all spectra with non-zero BI were visually inspected
in order to discard any in which the observed BAL trough was the
result of a poor reconstruction, or consisted solely of bad pixels.
Troughs that were visible in the data but were extended by a number of
bad pixels had their BI values recalculated with those pixels removed.
In cases where a small number of bad pixels appeared within a trough,
no changes were made. Again, the number of spectra where the automated
procedure requires tweaking is extremely small: 352 spectra had BAL
troughs modified or removed from the catalogue because of regions of
bad pixels, and 46 had troughs removed because of poor continuum fits.
Of the 46 spectra, the problem in all but 10 affected only the \mgii\
trough.  Cases in which a BAL trough exists but is not identified are
discussed in Sections \ref{sc:comparison} and \ref{sc:p_det}.

\section{The catalogue of BALQSOs}

\label{sc:results}

Data derived from the NMF fits to the SDSS DR6 quasar spectra are
presented in Table~\ref{tb:results}.  For each spectrum we present the
SDSS coordinate object name, right ascension (RA) and declination
(both using J2000 coordinates), the modified Julian date (MJD), plate
and fiber numbers of the spectroscopic observation, the S/N, flux and
luminosity at 1700\,\AA, redshift, $i$-magnitude, whether the object
was targeted in the SDSS as a HIZ or LOWZ quasar (see
Section~\ref{sc:selection}), whether the spectrum is the primary
spectrum for an object, and whether the object was used to generate
NMF components (GenComp).  Regarding the individual NMF fits, we
present the number of components used ($N_{\rm comp}$), whether that
number is a reduction from the number available (RedComp), the
$\chi_\nu^2$ of the fit, the number of pixels retained after masking
($N_{\rm pix}$), the index of the power-law slope fitted to identify
red objects (Slope), whether a slope correction was applied (SlCor),
and whether an additional mask was applied to cover a dip in the
reconstruction (DipMask) or defined manually (ManMask).

For each of \siiv\ $\lambda$1400, \civ\ $\lambda$1550,
\aliii\ $\lambda$1860 and \mgii\ $\lambda$2800 we list the BI, mean
depth (\md) and minimum and maximum velocities ($v_{\rm min}$, $v_{\rm
  max}$) of any identified BAL troughs, and the minimum and maximum
velocities covered by both the SDSS spectra and NMF components
($v_{\rm cov,min}$, $v_{\rm cov,max}$).  The coverage velocities are
set to zero when the BAL region has no coverage at all.  We also list
the number of bad pixels within the BI integration range ($N_{\rm
  BR}$) and within any BAL troughs ($N_{\rm BT}$), and the same for
pixels affected by sky lines ($N_{\rm SR}$ and $N_{\rm ST}$).
Finally, four flags are listed for each species: Inc is set if the
coverage of the BI region is incomplete, CBP denotes a manual change
has been made to the measured BI and associated properties because of
bad pixels in the originally-identified trough, CPF denotes a manual
change has been made because of a poor NMF fit, and BBP is set if
there is a broad region of bad pixels that could be concealing a BAL
trough.

The above information is presented for all quasars in the sample
described in Section~\ref{sc:sample}, regardless of whether any BAL
troughs are identified.

The SDSS names are taken from the SDSS DR7 Legacy Release where
available.  To calculate the S/N, flux and luminosity at
1700\,\AA\ the NMF reconstruction of the flux was smoothed by a median
filter with width 41 pixels, and the region between 1650 and
1750\,\AA\ was extracted.  The flux quoted ($F_{1700}$) is the median
value within this range, while the S/N ($SN_{1700}$) is the median of
the smoothed flux spectrum divided by the SDSS per-pixel noise
spectrum.  Bad pixels and those affected by sky lines were excluded
from the S/N calculation as their noise values are unreliable or
nonexistent, and the same pixels were excluded from the flux
measurement for consistency, particularly with regard to the
correlations measured in Section~\ref{ss:lum_snr}.

The luminosity
($\lambda L_{1700}$) is calculated from $F_{1700}$ using a luminosity
distance based on the redshift given in Table~\ref{tb:results}, and is
quoted after multiplying by $\lambda=1700$\,\AA\ for units of
erg\,s$^{-1}$.  The S/N, flux and luminosity are listed as zero if the range
1650--1750\,\AA\ is outside the rest-frame coverage of the SDSS
spectrum, or is covered entirely by bad pixels.  The $i$-magnitudes
are SDSS $i$-band PSF magnitudes, corrected for Galactic extinction
according to the dust maps of \citet{SFD98}.

The BI values are calculated according to equation~\ref{eq:bi} using
the rest-frame zero-velocity wavelengths 1402.77, 1550.77,
1862.79 and 2803.53\,\AA\ for \siiv, \civ, \aliii\ and \mgii,
respectively.  The corresponding wavelengths for
$v$$=$$-$25\,000\,\kms, the maximum velocity at which we search for
BAL troughs, are 1290.61, 1426.78, 1713.85 and 2579.37\,\AA,
respectively.  BAL systems with outflow velocities greater than this
limit are not identified, and in rare cases where a \civ\ outflow has
extremely high velocity it could potentially be misidentified as a
\siiv\ system.

The mean depth, \md, is the mean of
the flux ratio depth over all pixels in the BI-defined trough.  In
spectra with more than one distinct trough for a single species \md\ is
the mean over all troughs.  Similarly, the minimum and maximum
velocities listed give the minimum and maximum for all identified
absorption.

In calculating the BI and \md\ values, 12.0-\AA\ regions around
5578.5 and 6301.7\,\AA\ in the observed frame were interpolated
over to remove the effect of the prominent sky lines at those
wavelengths.  Although such an interpolation can smooth over regions
that genuinely rise above a flux ratio of 0.9, it is more common for
the presence of a sky line within a broad absorption trough to halt
the BI integration despite the true flux ratio still being below 0.9.
The redshift regions in which the sky lines fall within the \civ\ BI
integration region are 2.63$<$$z$$<$2.91 and 3.10$<$$z$$<$3.42.

Table~\ref{tb:results} includes data for both `primary' (best) and
`duplicate' observations, so many objects are listed more than once.  To
remove duplicates, leaving exactly one spectrum for each object in the
catalogue, only those spectra with Primary=1 (in column 14)
should be used.  Duplicate observations can be matched with their
corresponding primary observations using the SDSS object name (column
1) or coordinates (columns 2 and 3).

\subsection{Distributions of absorption properties}

In total, 3547 BALQSOs are identified, including 811 \siiv, 3296 \civ,
214 \aliii\ and 215 \mgii\ BALQSOs.  Note that many BALQSOs show
absorption in more than one species.  In these totals, and in all
following calculations, only `primary' SDSS spectra are used.
`Duplicate' spectra are discussed in Section~\ref{sc:duplicates}.

\begin{figure}
  \includegraphics[width=84mm]{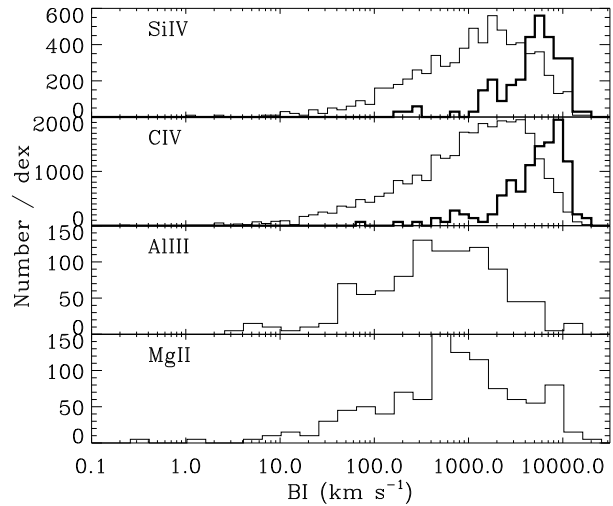}
  \caption{Distributions of BI values.  Thick lines in the upper two
    panels show the distributions for LoBALQSOs, i.e.\ those quasars
    that also have non-zero BI for \aliii\ or \mgii, scaled for
    clarity to have the same peak values as the overall distributions.
    Note that the \aliii\ and \mgii\ histograms have wider bins than
    the high-ionisation species due to the smaller number of quasars
    included.}
  \label{fg:bi.dist}
\end{figure}

The distributions of BI values for the four species examined are shown
in Fig.~\ref{fg:bi.dist}.  It can be seen that the two high-ionisation
species, \civ\ and \siiv, follow the same BI distribution, peaking at
2000--3000\,\kms\ in the log-space distribution, while the
low-ionisation species, \aliii\ and \mgii, peak below 1000\,\kms.
However, in cases where both a HiBAL and a LoBAL trough are visible,
the typical HiBAL BI is greater than for the HiBALQSO population as a
whole by a factor of $\sim$3, as shown by the thick lines in
Fig.~\ref{fg:bi.dist}.

\begin{figure}
  \includegraphics[width=84mm]{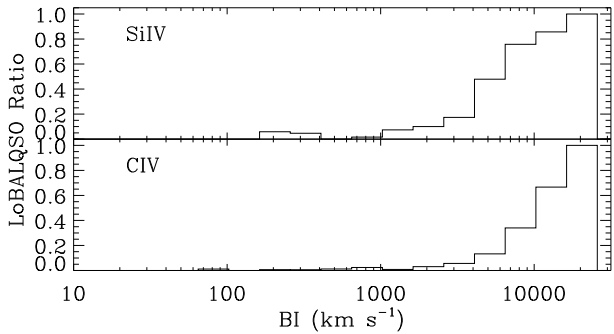}
  \caption{Ratio of LoBALQSOs to the total BALQSO population, as a
    function of \siiv\ (top panel) or \civ\ (bottom panel) BI.}
  \label{fg:bi.ratio}
\end{figure}

The extreme nature of the HiBAL absorption in LoBALQSOs is further
illustrated by Fig.~\ref{fg:bi.ratio}, in which the ratio of LoBALQSOs
to the total BALQSO population is shown as a function of \siiv\ or
\civ\ BI.  Only BALQSOs with complete coverage of the \aliii\ BAL
region were included in the calculation.  The LoBALQSO population
consists almost entirely of objects with HiBAL BI values of several
thousand \kms\ or more, and conversely almost all BALQSOs with HiBAL
BI$>$10000\,\kms\ also exhibit LoBAL absorption.

\begin{figure}
  \includegraphics[width=84mm]{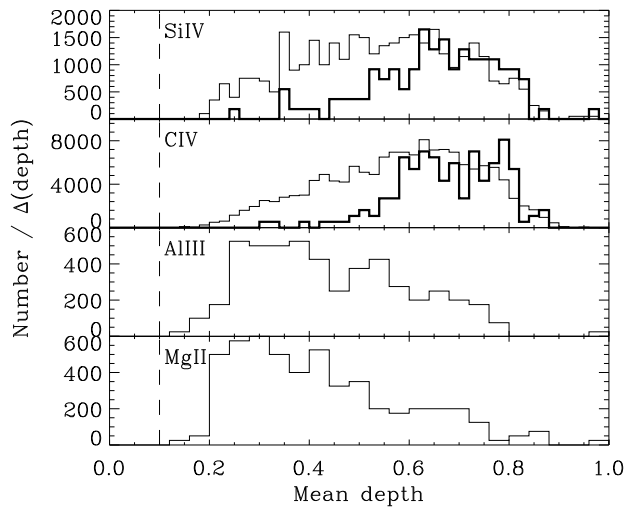}
  \caption{As Fig.~\ref{fg:bi.dist}, but for mean depth values.  The
    dashed line marks the minimum mean depth of 0.1 allowed by the BI
    definition (equation~\ref{eq:bi}).}
  \label{fg:md.dist}
\end{figure}

\begin{figure}
  \includegraphics[width=84mm]{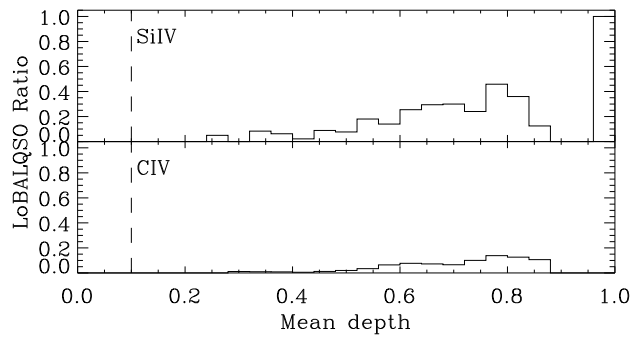}
  \caption{As Fig.~\ref{fg:bi.ratio}, but for mean depth values.  The
    dashed line marks the minimum mean depth of 0.1 allowed by the BI
    definition (equation~\ref{eq:bi}).}
  \label{fg:md.ratio}
\end{figure}

The differences between high- and low-ionisation absorption properties
can also be seen in the distributions of \md\ values, shown in
Fig.~\ref{fg:md.dist}: the high-ionisation species peak at
$\md$$\simeq$0.6, while the low-ionisation species peak at
$\simeq$0.3.  As expected from the BI values, the typical depth of a
HiBAL trough in a LoBALQSO is greater than in the general HiBALQSO
population.  However, Fig.~\ref{fg:md.ratio} shows there is a large
population of BALQSOs with very deep HiBAL troughs but no LoBAL
absorption.

\subsection{Repeat observations}

\label{sc:duplicates}

A number of quasars were observed more than once in the SDSS spectroscopic
survey.  Such quasars provide a convenient test of the reliability of
the NMF reconstructions and the resulting balnicity measurements, as
the actual balnicities of most objects are not expected to vary greatly over the
time-scales of the survey \citep{Lundgren07,Gibson08}.

\begin{figure}
  \includegraphics[width=84mm]{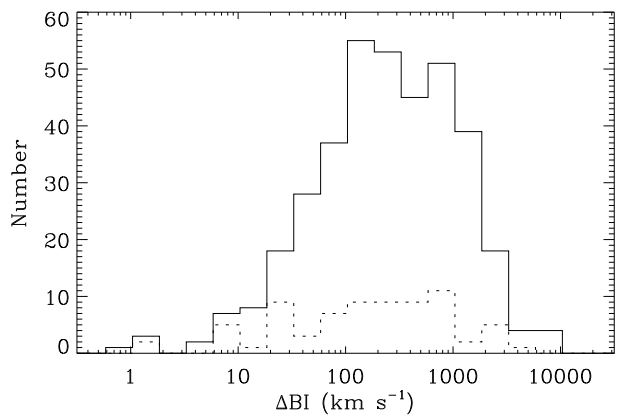}
  \caption{Solid line: distribution of differences in \civ\ BI values between
    different observations of the same quasars.  Dotted line:
    distribution of \civ\ BI values for which the paired observation
    has ${\rm BI} = 0$.}
  \label{fg:duplicates}
\end{figure}

Counting only those observations for which the \civ\ Inc and \civ\ BBP
flags are not set, there are 2051 quasars with between two and seven
observations in DR6.  Of the 3288 pairs of observations available,
there are 300 for which both NMF reconstructions give a \civ\ BALQSO
classification, and a further 73 for which one reconstruction does but
not the other.  The distribution of differences between the pairs of
\civ\ BI values, $\Delta$BI, is shown in Fig.~\ref{fg:duplicates}.
For the majority of cases the discrepancy is less than 1000\,\kms.
Visual inspection confirms that, as expected, most of the pairs with
larger $\Delta$BI consist of observations with significantly differing
S/Ns.  As detailed in Section~\ref{sc:p_det}, the measurement of
BALQSO properties is very sensitive to S/N, even in the ideal
situation in which the NMF reconstructions do not vary.

The presence of pairs in which one observation is classified as a
BALQSO, but the other is not, is confirmation that not all BAL troughs
are identified by the NMF procedure.  The resulting incompleteness is
addressed in Section~\ref{sc:p_det}.

\subsection{Observed BALQSO fractions}

\begin{figure}
  \includegraphics[width=84mm]{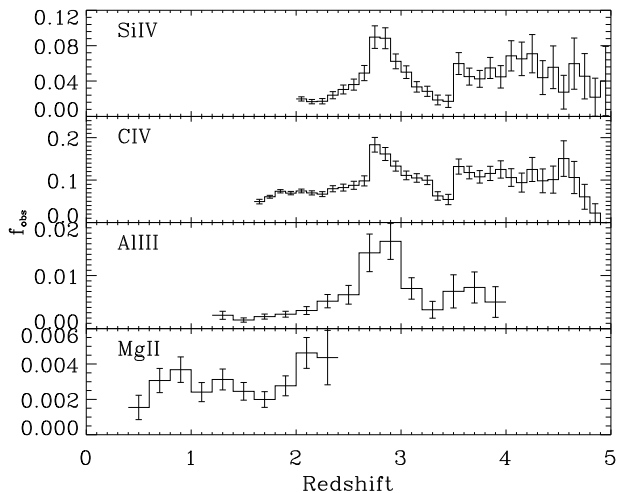}
  \caption{Observed BALQSO fractions as a function of redshift.  As in
    Figs. \ref{fg:bi.dist} and \ref{fg:md.dist} different bin widths
    have been used for different species.  Binomial errors are shown.
    The different redshift ranges covered reflect the redshifts at
    which each emission line is observable in the SDSS spectra.}
  \label{fg:f_obs.z}
\end{figure}

The observed BALQSO fractions for \siiv, \civ, \aliii\ and \mgii\ as a
function of redshift are shown in Fig.~\ref{fg:f_obs.z}.  Only spectra
for which the relevant Inc and BBP flags were not set were included
when calculating these fractions.  There is a strong apparent redshift
dependence of the BALQSO fractions, the implications of which are
addressed in Sections \ref{sc:p_det}--\ref{sc:intrinsic}.  The
highest-redshift BALQSO identified is a \civ\ BALQSO at redshift,
$z$$=$5.0314, but for $z$$>$5 no \civ\ BALQSO fraction is measured
because all spectra have incomplete coverage of the \civ\ absorption
region.  Of the 47 spectra at $z$$>$5 for which complete coverage of
the \siiv\ absorption region is available, none are identified as
\siiv\ BALQSOs.  Summing over all redshifts, the BALQSO fractions for
\siiv, \civ, \aliii\ and \mgii\ are 3.4$\pm$0.1, 8.0$\pm$0.1,
0.38$\pm$0.03 and 0.29$\pm$0.02 per cent, respectively.  The
uncertainties quoted are 1-$\sigma$ binomial errors.

\begin{figure}
  \includegraphics[width=84mm]{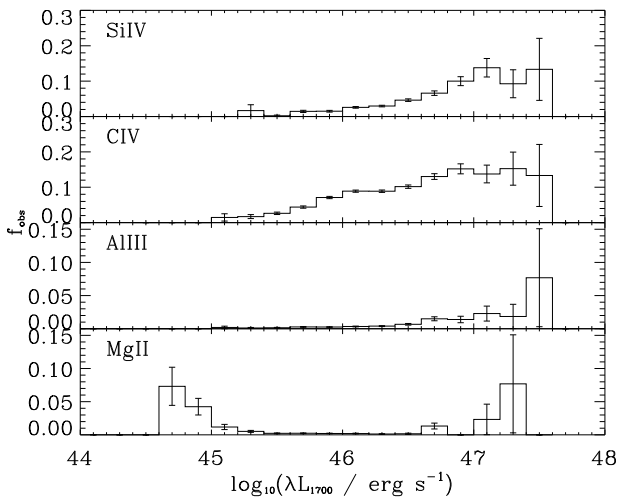}
  \caption{As Fig.~\ref{fg:f_obs.z}, but with the BALQSO fractions as
    a function of luminosity.}
  \label{fg:f_obs.l}
\end{figure}

The observed fractions as a function of $\lambda L_{1700}$ are shown in
Fig.~\ref{fg:f_obs.l}. Only objects with redshifts
1.2$\lesssim$$z$$\lesssim$4.6, i.e. where 1700\,\AA\ rest-frame is
present in the SDSS spectra, are included.  The luminosities are not
corrected for dust extinction at this point.  The distributions for
\siiv, \civ\ and \aliii\ all show a strong tendency towards higher
BALQSO fractions at high luminosity, but this result is strongly
biased by the relative probabilities of detecting a given BAL trough
at different S/Ns; this bias is addressed in
Section~\ref{sc:p_det}.  The fraction of \mgii\ BALQSOs shows a marked
increase at $\lambda L_{1700} < 10^{45}$\,erg\,s$^{-1}$ that is not observed in
the other species; the BALQSOs contributing at these luminosities are
all very heavily dust-reddened, resulting in a very low flux at
1700\,\AA.

\section{Comparison to previous results}

\label{sc:comparison}

G09 presented a catalogue of BALQSOs in the SDSS DR5, in
which the unabsorbed emission was estimated by fitting a
dust-reddened power law with Voigt profiles for the strong emission
lines.  The DR5 spectra are a large subsample of the DR6 spectra used
here and a comparison of the results is thus possible.

\begin{figure}
  \includegraphics[width=84mm]{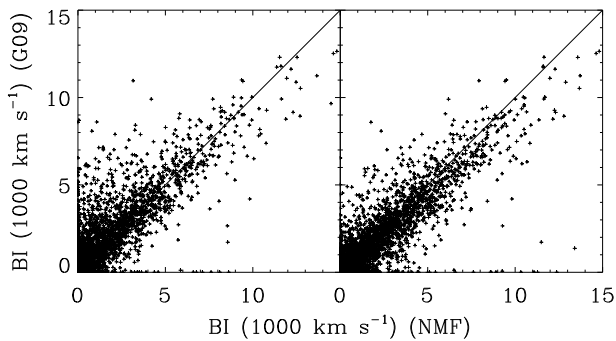}
  \caption{Comparison of \civ\ BI values presented in this paper and in
    \citet{Gibson09}.  Left panel: NMF-derived BI values as presented in
    Section~\ref{sc:results}.  Right panel: NMF-derived BI values after
    boxcar smoothing the observed flux.}
  \label{fg:g09.bi}
\end{figure}

The left panel of Fig.~\ref{fg:g09.bi} compares the \civ\ BI values
presented in this paper to those in G09, including only those spectra
that were available in DR5.  It is immediately clear that in general
there is good agreement between the measurements.  However, there are some
notable differences.  In particular,
there are many more objects for which the G09 BI value is several
thousand \kms\ higher than the NMF-derived BI value than there are for
which the reverse is true; this can be seen in the asymmetry in the
distribution of points around the line of equality.  Similarly, there
are 1206 objects that are identified as \civ\ BALQSOs in G09 but not
in this paper, but only 149 that are identified as \civ\ BALQSOs here
but not in G09.

Before calculating the BI values, G09 smoothed the observed flux of
each object using a boxcar with a width of three pixels.  This smoothing
was performed to reduce the number of cases where the noise in a
single pixel breaks up a BAL trough.  When the same smoothing is applied
before recalculating the NMF-derived BI values, the results from the
two catalogues follow each other more closely, as shown in the right
panel of Fig.~\ref{fg:g09.bi}.  The results presented in
Section~\ref{sc:results} use the un-smoothed spectra; the resulting
incompleteness is corrected for in the following Sections.  Visual
inspection confirms that the spectra for which the BAL status is
different in different catalogues tend to be marginal cases for which
it is debatable whether any absorption present is deep or broad enough
to be classified as a `broad absorber'.

For the strongest absorbers (BI $\gtrsim 10\,000$\,\kms)
Fig.~\ref{fg:g09.bi} shows a trend for the NMF-derived BI values to
be greater than those from G09.  The reason for the discrepancy is not
always clear but the trend implies that, for high-BI quasars, the NMF
algorithm tends to place the reconstructed continuum higher than G09
reconstructions.

\begin{figure}
  \includegraphics[width=84mm]{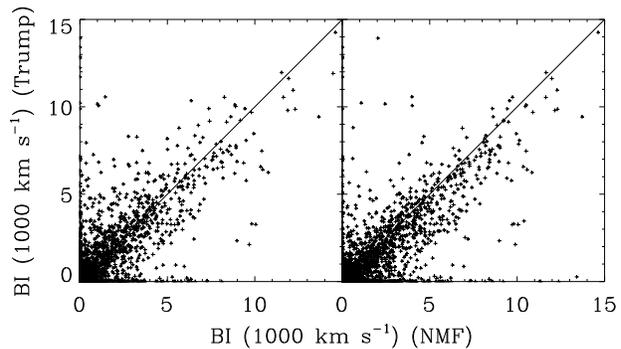}
  \caption{As Fig.~\ref{fg:g09.bi}, but comparing the NMF results to
    those from \citet{Trump06}.}
  \label{fg:trump.bi}
\end{figure}

Similar trends can be seen when comparing to the DR3 catalogue
presented by \citet{Trump06}, as shown in Fig.~\ref{fg:trump.bi},
although the initial asymmetry is less pronounced and there is a
tendency for the smoothed NMF-derived BI values to be higher than the
values from \citet{Trump06} across the entire range, again implying
the NMF reconstructions have a higher continuum level.

The varying classifications between these three catalogues, each of
which used the same metric on essentially the same data, emphasizes
the importance of an accurate fit to the unabsorbed continuum.
Relatively small errors in the position of the continuum can often
change a classification between non-BAL and BALQSO, or change the
measured BI by thousands of \kms.  We expect that the NMF-derived fits
presented in this paper will be consistently more accurate than the
fits used to create previous catalogues, due to the manner in which
the NMF procedure combines data from the entire observed spectrum to
reconstruct the continuum.  In cases in which large portions of the
spectrum are absorbed, such a method has clear advantages over one
that attempts to directly fit line profiles in the absorbed regions.
Visual inspection of reconstructions of synthetic BALQSO spectra, such
as those described in Section~\ref{sc:p_det}, suggests that the NMF
procedure reliably produces accurate reconstructions of quasar
continua, even in cases where large regions are affected by
absorption.

Recently \citet{Scaringi09} have presented a BALQSO catalogue in which
`learning vector quantisation' (LVQ), a machine-learning algorithm,
was used to classify quasars as BALQSOs or non-BAL quasars.  These
classifications were compared to those from G09 and, in the cases
where the two disagreed, a visual inspection was performed to provide
a final hybrid-LVQ classification.  

The final hybrid-LVQ catalogue leaves out very few of the DR5 quasars
that are classified as BALQSOs in this paper.  Conversely, nearly 1400
quasars have BI=0\,\kms\ when measured from the NMF results but are
classified as BALQSOs by \citet{Scaringi09}.  In part this discrepancy
is due to the increased probability of the G09 fits producing a
non-zero BI relative to the NMF results, but the decision by
\citet{Scaringi09} to include absorbers with velocity $<$3000\,\kms\ or
velocity width $<$2000\,\kms\ if a visual inspection indicates they have
similar properties to other BALQSOs, results in the inclusion of many
objects that by definition will always be excluded from a purely
BI-based catalogue.  The resulting discrepancy serves to underline
the current difficulties in objectively defining `broad absorption'.

The public release of the \civ-region NMF-continuum fits, for each of
the 48\,146 spectra in the sample with complete or partial coverage of
the \civ\ region, allows anyone to use them to define BALQSO samples
using any metric they wish.  Additionally, future continuum fits to
the same objects can be compared to the NMF-derived fits to aid in
identification and explanation of discrepancies.

\section{C\,{\sevensize\bf IV} BALQSO detection efficiency}

\label{sc:p_det}

The catalogue presented in Section~\ref{sc:results} is not
complete.  A number of BALQSOs will exist in the SDSS DR6
spectroscopic survey but have a measured BI of zero.  There are two
principal reasons why this can occur.  Firstly, the reconstruction of
the emission may place the continuum level too low, reducing the
apparent depth of a trough or eliminating it completely.  Secondly,
even for a perfect reconstruction, a noise spike in the observed
spectrum may lie above 90 per cent of the unabsorbed emission, halting
the BI integration.  This incompleteness in identifying BALQSOs from
observed spectra has previously been noted (\citealt{Knigge08}; G09)
but not quantified.  Because the level of
incompleteness is strongly dependent on the S/N of the observed
spectra, an extensive analysis is required in order to quantify the
BALQSO fraction within the SDSS, $f_{\rm SDSS}$, as a function of any
parameters that are correlated with spectrum S/N, such as luminosity
or redshift.

It is also possible for non-BAL quasars to be incorrectly identified
as BALQSOs due to a reconstruction that places the continuum too
high. However, following the results of the comparisons in
Section~\ref{sc:comparison}, such false positives are expected to be
rare in the catalogue presented here.

\begin{figure}
  \includegraphics[width=84mm]{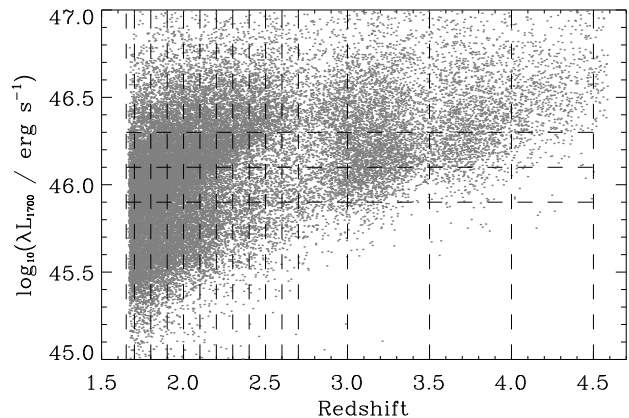}
  \caption{Redshift and luminosity bins for which the BALQSO detection
    fraction was calculated.  Grey points show the redshift and
    luminosity of all quasars in DR6 with neither the \civ\ Inc nor
    the \civ\ BBP flags set; dashed lines show the boundaries of
    the bins.}
  \label{fg:pdetbins}
\end{figure}

The probability of a particular \civ\ BALQSO being detected by the NMF-based
routine, $p_{\rm det}$, was estimated by inserting BAL troughs with known
properties into non-BAL quasar spectra and processing the resulting
synthetic BALQSOs in the same way as for observed spectra.  The value
of $p_{\rm det}$ for any selected set of quasar and BAL properties is then
simply the number of correctly identified BALQSOs, divided by the
total number of synthetic spectra created.  As $p_{\rm det}$ was expected
to be a strong function of redshift and S/N it was calculated
separately for each of the redshift--luminosity bins shown in
Fig.~\ref{fg:pdetbins}.

The methods described below could be applied to BAL systems in any
species, but in this work we examine only \civ\ systems, by far the
most common form of BALQSO.

\subsection{Synthetic BALQSO spectra}

\begin{figure}
  \includegraphics[width=84mm]{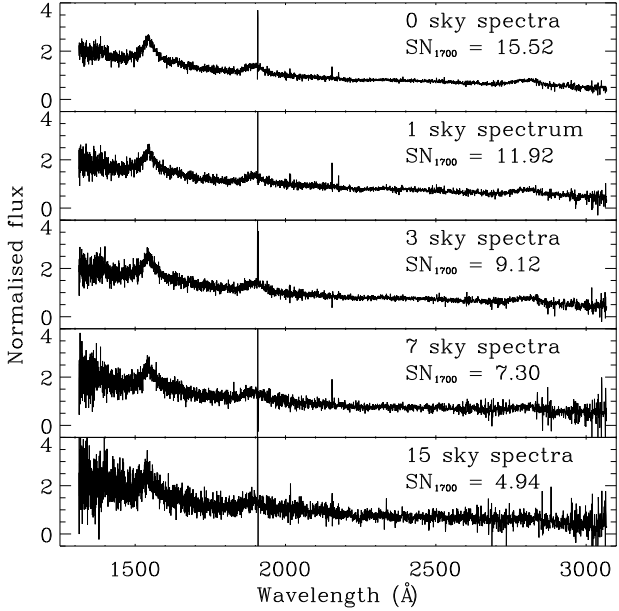}
  \caption{Top panel: Flux spectrum of SDSS J133533.69+032610.8.
    Other panels: the same spectrum after addition of 1, 3, 7 and 15
    sky spectra.}
  \label{fg:addsky}
\end{figure}

In each redshift bin a sample of 50 quasars was chosen such that each
quasar had 9$\le${\tt SN\_R}$<$25 and no significant absorption in the
\civ\ trough region.  For the 4.0$\le$$z$$<$4.5 redshift bin the
minimum {\tt SN\_R} was reduced to 7 to ensure sufficient spectra were
available.  To create spectra with the same properties but over a
range of S/N, the observed quasar spectra were degraded by adding in sky
spectra from the same SDSS plate. An example of
this degradation is shown in Fig.~\ref{fg:addsky}, where a quasar
spectrum has 1, 3, 7 and 15 sky spectra added to reduce the S/Ns by
nominal factors of $\sqrt{2}$, $\sqrt{4}$, $\sqrt{8}$ and $\sqrt{16}$,
although the actual S/N was always recalculated for each degraded
spectrum.

The spectra were degraded by increasing factors of $\sqrt{2}$ in S/N
until there were insufficient sky spectra from the SDSS plate to
continue; in most cases this halted the process at 31 additional sky
spectra, for a S/N nominally reduced by a factor of
$\sqrt{32}\approx5.66$.

\begin{figure}
  \includegraphics[width=84mm]{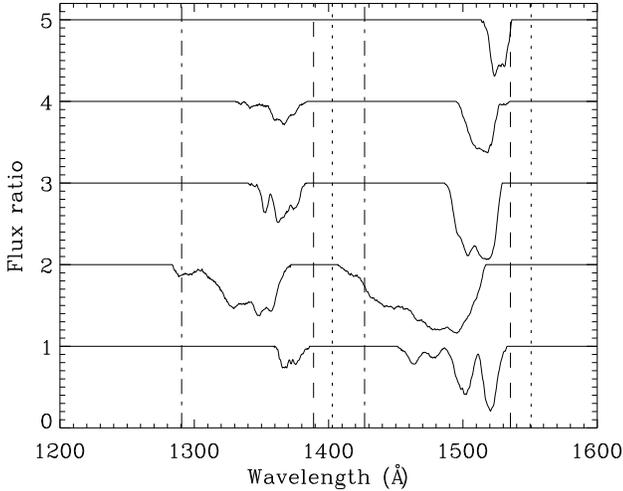}
 \caption{Flux ratios used to generate synthetic BALQSOs to test the
   BALQSO identification completeness.  Each profile is offset for
   clarity.  The short-dashed lines mark the
   zero-velocity wavelengths of \siiv\ and \civ, and the long-dashed
   and dot-dashed lines mark the wavelengths at $v=-3000$ and
   $-25000$\,\kms, respectively.  BI values for \siiv/\civ\ are, from
   top to bottom, 0/613, 272/1548, 1246/4231, 4434/9367 and 152/1327
   (all values in \kms).}
 \label{fg:fr.btdl.profiles}
\end{figure}

The BAL troughs used to create synthetic BALQSOs were created from
the flux ratio profiles of a set of five actual BALQSOs, chosen to
provide a range of BAL velocity widths.  The initial flux ratios were
generated by dividing the observed spectrum by the NMF reconstruction,
and smoothing the result by a boxcar window with width 15 pixels.
The maximum flux ratio was set to unity, and any absorption
outside the principal \civ\ and \siiv\ troughs was erased.  The
resulting smoothed ratios are shown in Fig.~\ref{fg:fr.btdl.profiles}.

\begin{figure}
  \includegraphics[width=84mm]{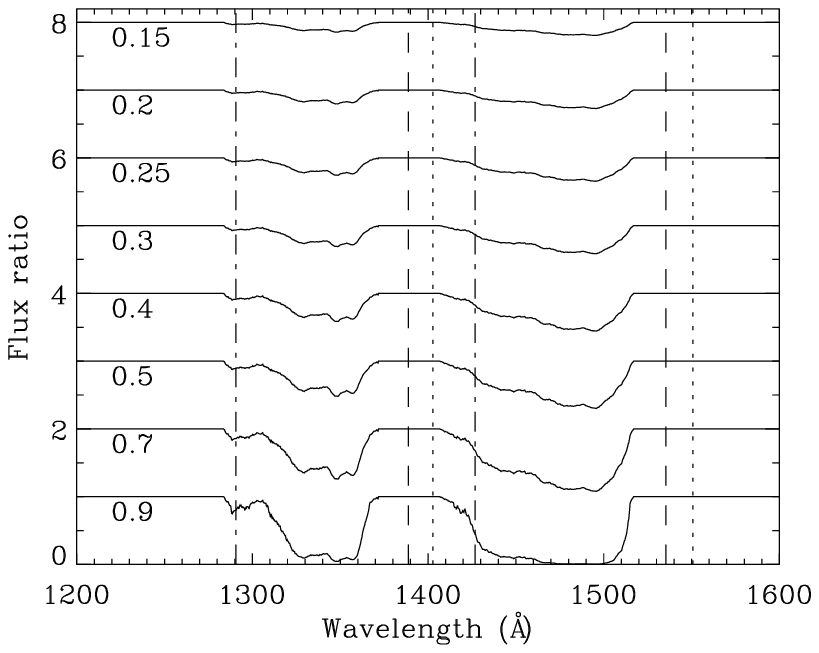}
  \caption{Final flux ratio profiles with \civ\ \md\ values as shown,
    generated from the 4th profile from the top in
    Fig.~\ref{fg:fr.btdl.profiles}.  Each profile is offset for
   clarity.  The short-dashed lines mark the
   zero-velocity wavelengths of \siiv\ and \civ, and the long-dashed
   and dot-dashed lines mark the wavelengths at $v=-3000$ and
   $-25000$\,\kms, respectively.  BI values for \siiv/\civ\ are, from
   top to bottom, 142/800, 555/1847, 1020/2820, 1475/3747, 2469/5625,
   3450/7501, 5655/11198 and 8775/14593 (all values in \kms).}
  \label{fg:fr.btdl.depth}
\end{figure}

Preliminary tests suggested that the fraction of BAL troughs recovered
depended more strongly on the mean depth, \md\ than on the BI itself,
so each of the flux ratio profiles in Fig.~\ref{fg:fr.btdl.profiles}
was scaled to create profiles with \civ\ $\md = 0.15$, 0.2, 0.25, 0.3,
0.4, 0.5, 0.7 and 0.9.  A simple linear scaling was used.
Fig.~\ref{fg:fr.btdl.depth} shows the resulting troughs at different
values of \md\ for one of the input flux ratio profiles.

\begin{figure}
  \includegraphics[width=84mm]{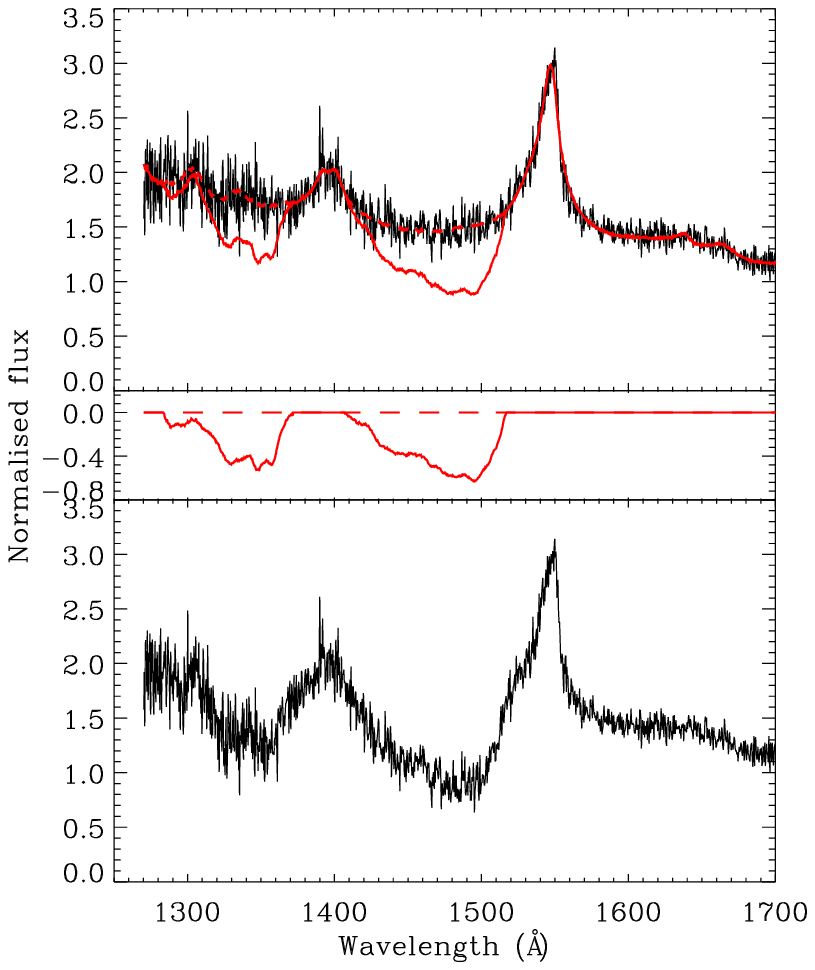}
  \caption{Method of inserting a BALQSO trough into a non-BAL quasar
    spectrum.  Top panel: Flux spectrum (black) of SDSS
    J122626.22+270437.0 with the initial NMF
    reconstruction (dashed red) and the reconstruction multiplied by
    the chosen flux ratio spectrum (solid red).  Middle panel:
    difference between the initial reconstruction and the reconstruction
    multiplied by the flux ratio spectrum.  Bottom panel: quasar
    spectrum after subtracting the reconstruction difference spectrum
    shown in the middle panel.}
 \label{fg:synbal}
\end{figure}

To insert the troughs into non-BAL quasar spectra, simply multiplying the
reference spectra by the selected flux ratio profiles would have
artificially reduced the noise within the troughs.  Instead the
original NMF reconstructions of the non-BAL quasar spectra were
multiplied by the flux ratios, and the resulting trough shapes -- the
difference between the reconstruction and the reconstruction
multiplied by the flux ratio -- were subtracted from the observed
spectra, as shown in Fig.~\ref{fg:synbal}.

By inserting each synthetic BAL profile at each mean depth into each
of the spectra in the test sample for all levels of S/N degradation,
the 50 input spectra in each redshift bin were expanded to over
10\,000, covering extended ranges of S/N and BAL properties.  NMF
reconstructions of these synthetic BALQSO spectra were calculated by
the method described in Section~\ref{sc:reconstructions} to determine
if the BAL troughs would be detected.

\subsection{Relating luminosity to signal-to-noise ratio}

\label{ss:lum_snr}

The variation in $p_{\rm det}$ with S/N is crucial because S/N correlates
with important physical parameters such as redshift and luminosity.
The synthetic BALQSO spectra have known redshifts, from the original
non-BAL quasars on which they are based, but the luminosities of the
degraded spectra are not well-defined. To relate the S/N to luminosity,
the $SN_{1700}$ and $\lambda L_{1700}$ values for
all quasars with $SN_{1700} < 25$ were extracted from the catalogue.
In each redshift range a linear regression fit was made to empirically
determine the relationship between S/N and luminosity.  The values
used to define the luminosity bins were then converted to S/N values
using these linear relationships, allowing each synthetic spectrum to
be assigned to a redshift--luminosity bin.

\subsection{Detection probabilities}

\begin{figure*}
  \includegraphics[width=168mm]{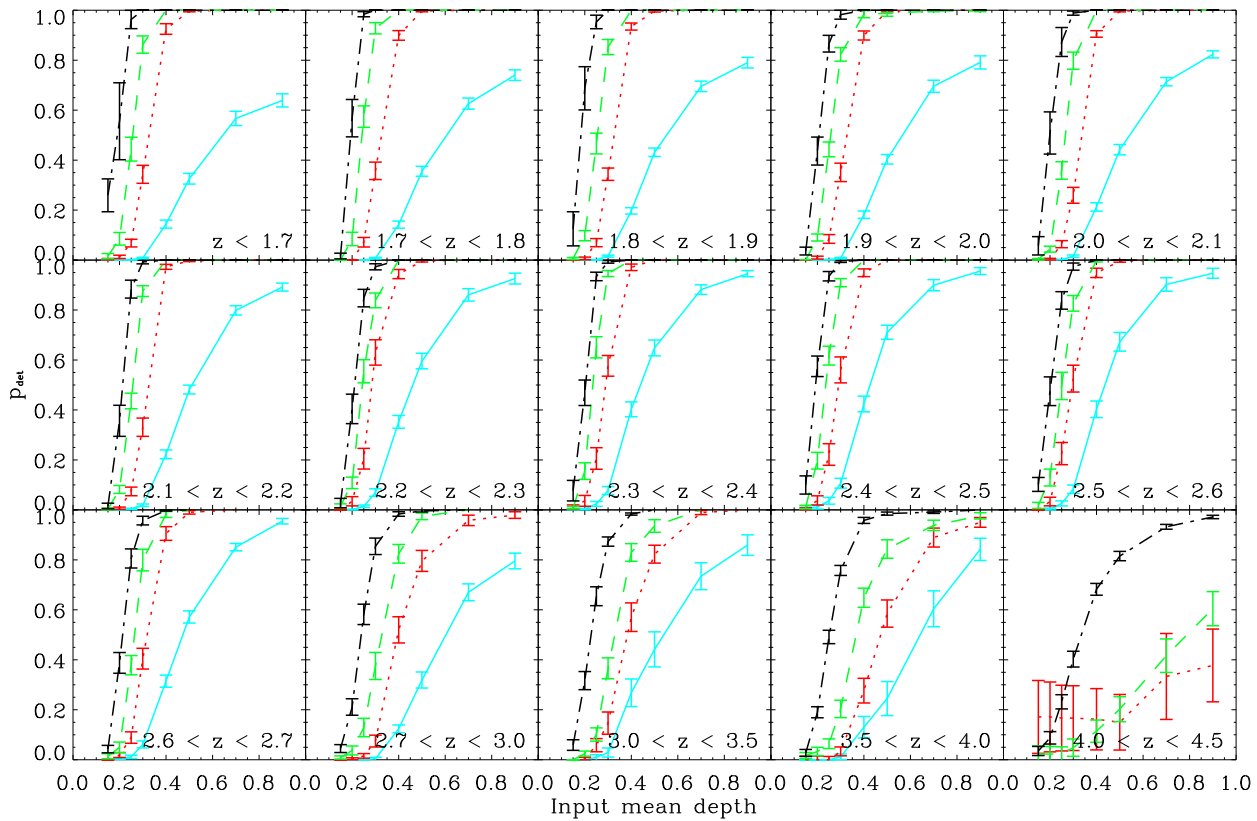}
  \caption{Probability of the NMF routine detecting a BALQSO in SDSS
    as a function of input \md.  Each panel represents a single
    redshift bin, as labelled.  The curves within each panel represent
    different luminosities: $\lambda L_{1700} <
    10^{45.9}$ (solid blue), $10^{45.9} \le
    \lambda L_{1700} < 10^{46.1}$ (dotted red),
    $10^{46.1} \le \lambda L_{1700} <
    10^{46.3}$ (dashed green) and $10^{46.3}
    \le \lambda L_{1700}$ (dot-dashed black) 
    (all luminosities in erg\,s$^{-1}$).}
  \label{fg:p_det}
\end{figure*}

The derived values of $p_{\rm det}$ as a function of input mean depth,
$\md_{\rm in}$, are shown in Fig.~\ref{fg:p_det} for all redshift and
luminosity bins.  The uncertainties were calculated from bootstrap
realisations, in each of which 50 spectra were chosen at random with
replacement from the 50 input spectra.

As expected, the detection probabilities are highest for quasars with
high luminosity (and hence high S/N) and deep BAL troughs.  In
general, for $\lambda L_{1700} \ge 10^{45.9}$\,erg\,s$^{-1}$, there is a value
of $\md_{\rm in}$ above which all or nearly all BAL troughs are detected; below
this depth $p_{\rm det}$ falls rapidly.  In the redshift range
4.0$\le$$z$$<$4.5 (bottom-right panel in Fig.~\ref{fg:p_det}) none of
the synthetic spectra reached the low S/N required for the lowest
luminosity bin, although the results from the higher-luminosity bins
suggest $p_{\rm det}$ would be very low.  With only four quasars observed
in this bin, we exclude the bin in the following analysis.

The detection probabilities calculated here were based on the actual
mean depth of the BAL trough, but an examination of the results from
the synthetic BALQSOs suggested that on average the observed mean
depth, $\md_{\rm obs}$, was slightly deeper than the input mean depth,
$\md_{\rm in}$.  The offset is due in part to (i) a tendency for the
shallow wings of the input absorption profiles to be excluded from the
observed BI regions and (ii) the possibility for an observed trough,
scattered to smaller depths, to be mis-identified as a non-BAL quasar.
To correct for this trend, in each redshift--luminosity bin a linear
regression line was fitted to $\md_{\rm in}$ as a function of
$\md_{\rm obs}$, and the observed mean depths presented in
Section~\ref{sc:results} were reduced according to the regression line
before $f_{\rm SDSS}$ was calculated.  The typical corrections are
small, with a median $\Delta\md$ of only 0.03.  The corrected
depth, $\md_{\rm cor}=\md_{\rm obs}-\Delta\md$, is taken to be
equivalent to the input depth, $\md_{\rm in}$, of the synthetic
quasars.

\section{Differential SDSS target selection}

\label{sc:selection}

Any BALQSO fraction derived directly from the quasar spectra in the
SDSS will not be the intrinsic BALQSO fraction, because the SDSS
spectroscopic survey is not 100 per cent complete.  Quasar candidates
were chosen for spectroscopic observations based on their photometric
properties, as well as specifically targeting sources identified in
the FIRST radio catalogues \citep{BWH95}, giving an overall
completeness, for quasars brighter than an {\it observed} i-band
magnitude, of $>$90 per cent \citep{Richards02}.  Crucially,
the presence of material causing a BAL trough along the line-of-sight
to a quasar changes the quasar's observed photometric properties.  As
a result, the probability of a BALQSO being selected by the target
selection algorithm is different to that for a non-BAL quasar whose
properties are otherwise the same.  Differential selection
probabilities were calculated by \citet{Reichard03b}, using simulated
BALQSO and non-BAL quasar magnitudes; we follow a similar but more 
involved procedure here.

First, model quasar spectra are generated covering a wide range of
expected quasar properties (Section~\ref{ss:synspec}).  As well as
different BAL properties we simulate different levels of dust
reddening, a range of Lyman limit cut-off wavelengths and different
continuum power laws.  Each of these factors has an effect on the
photometric properties of quasars and hence can change the SDSS
completeness.  In Section~\ref{ss:synmag} we calculate observed
$ugriz$ magnitudes from the model spectra at a range of redshifts and
luminosities, allowing us to process them with the SDSS quasar target
selection algorithm to determine the probability each model quasar
would be targeted.  The resulting completeness values are presented as
contour plots in redshift--magnitude space in
Section~\ref{ss:contours}.

In order to make use of the completeness values derived for each
individual model quasar, they must be weighted according to the
expected probability distribution functions of each input parameter.
In Section~\ref{ss:distributions} we describe our methods for
determining such distributions.  Finally, in
Section~\ref{ss:completeness} we combine the weighted completeness
values with an input luminosity distribution to derive the overall
completeness for BALQSOs and non-BAL quasars as a function of redshift
and luminosity.

\subsection{Model quasar spectra}

\label{ss:synspec}

The wavelength coverage required to synthesise SDSS $ugriz$
photometry is greater than that provided by the SDSS spectra, so more
extended model spectra must be used. The reference quasar model for the
investigation is that described in \citet{Maddox08}. The Maddox et al.\
model quasar SED reproduces the variations of the median colours of the
SDSS quasars over the full redshift range 0.2$<$$z$$<$5.0 to a high 
degree of accuracy. A graphical indication of the success of an earlier
version of the model in reproducing the colours of SDSS quasars can be
found as fig.~8 of \citet{Chiu07}. The implementation employed here
incorporates a refined parametrization of the Ly-$\alpha$ forest
opacity based on the work of \citet{FG08}. The model
results in an excellent match to the observed median $ugr$ colours of
the SDSS quasars over the redshift range 1.7$<$$z$$<$5.0, representing a
significant improvement over the parametrization based on the results
of \citet{Songaila04} that was employed previously.

The \citet{Maddox08} model SED does reproduce very well the locus of
observed quasar colours over an extended range of redshifts. However,
it uses a fixed continuum model while the quasar population exhibits
an intrinsic range in overall continuum shape.  As only a small
fraction of quasars deviate significantly from the model SED such
quasars typically have little effect on the overall SDSS completeness.
None the less, there are regions of parameter space in which the quasar
locus passes through the colour-volume occupied by main sequence stars
and the predicted completeness values plummet, while those quasars
with unusually blue SEDs skirt the stellar main sequence and do
satisfy the SDSS quasar selection criteria.  In such regions the
`blue' quasars do make a significant contribution to the completeness
values.  To include this effect an additional series of model quasars,
with the rest-frame ultraviolet power law slope $\alpha$ ($f_\nu =
\nu^{-\alpha}$) bluer by 0.5, was generated.

The reference quasar spectra were modified in the following ways to
reproduce the range of SEDs present among the quasar population.  The
spectrum was dust-reddened using an empirically-derived extinction
curve (Maddox et al.\ 2010, in preparation) with \ebv\ values of 0.0,
0.05, 0.1, 0.15 and 0.2. The Maddox et al.\ (2010) extinction curve is
very similar to the extinction curve of the Small Magellanic Cloud
(SMC), which is frequently used in studies of quasars. Like the
SMC curve, there is no 2175-\AA \ feature but the increase in
extinction below 1600\,\AA \ is somewhat shallower, although still
always greater than for a Large Magellanic Cloud extinction curve.

Changing the wavelength of the Lyman limit cut-off, $\lambda_{\rm
  LL}$, can have a large effect on the observed quasar colours,
particularly by changing the $u$-band flux and hence the $u$--$g$
colour.  Such an effect can move quasars into or out of the stellar
locus, changing their targeting status with only a small shift in
$\lambda_{\rm LL}$, a feature which \citet*{Prochaska09} and
\citet{WP10} have highlighted recently.  To explore the impact of a
Lyman limit cut-off we generated sets of objects following the tracks
in $ugriz$ magnitude space defined by varying $\lambda_{\rm LL}$
between 600 and 912\,\AA.  The objects were spaced along the tracks
such that the step size, $\Delta m = \sqrt{{\Delta u}^2 + {\Delta g}^2
  + {\Delta r}^2 + {\Delta i}^2 + {\Delta z}^2}$, was constant, with
the number of steps chosen individually for each test quasar to give
$\Delta m \simeq 0.1$.

BALQSOs were simulated by employing a modified subset of the flux
ratios used in Section~\ref{sc:p_det}.  The flux ratio profiles used
were the first, third and fourth from the top in
Fig.~\ref{fg:fr.btdl.profiles}, each with mean depths of 0.15, 0.3,
0.5, 0.7 and 0.9. In each case the \civ\ trough profile was replicated
at the wavelengths of the \siiv, \nv\ and Ly$\alpha$ lines. In the
case of the \siiv\ and \nv\ lines the \civ\ trough was reduced to 80
per cent of its original depth, in order to ensure consistency between
different profiles representing different BI ranges.  In addition, the
flux bluewards of 1050\,\AA\ was reduced by the mean depth of the
\civ\ trough, to approximate the combined effect of broad absorption
from a number of other high-ionisation species.  The existence of
these troughs can be seen directly by comparing composite SDSS spectra
of BALQSOs and non-BAL quasars with redshifts 3.0$<$$z$$<$4.0 (such as
those in Fig.~\ref{fg:ebv_composites}), where the rest-frame
800--1100\,\AA\ wavelength region is visible in the spectra, showing a
systematic depression in the BALQSOs compared to the non-BAL quasars
below $\simeq$1050\,\AA.

The BALQSO completeness values calculated below were interpolated to
cover all eight of the mean depths for which the detection efficiency
was calculated in Section~\ref{sc:p_det}, for each of the three
profiles used, giving $8\times3=24$ BI/\md\ ranges.

The recipe adopted for the BAL trough simulations closely matches the
observed properties of BAL quasars among the SDSS spectra.  The
non-BAL quasar spectra were also retained and the full suite of model
spectra covers an extended range in the reddening and BAL properties
of quasars.

\subsection{Synthetic SDSS magnitudes}

\label{ss:synmag}

In order to measure the completeness for the model quasar spectra,
SDSS magnitudes were generated for each and multiple realisations were
created by scattering according to typical photometric errors.  The
sets of scattered magnitudes were processed by the SDSS quasar target
selection algorithm to determine if each would be targeted.  The
completeness for each model quasar could then be estimated as the
fraction of realisations, scattered from the true magnitudes for that
model, that were targeted.  The procedure is described in detail below.

\begin{figure}
  \includegraphics[width=84mm]{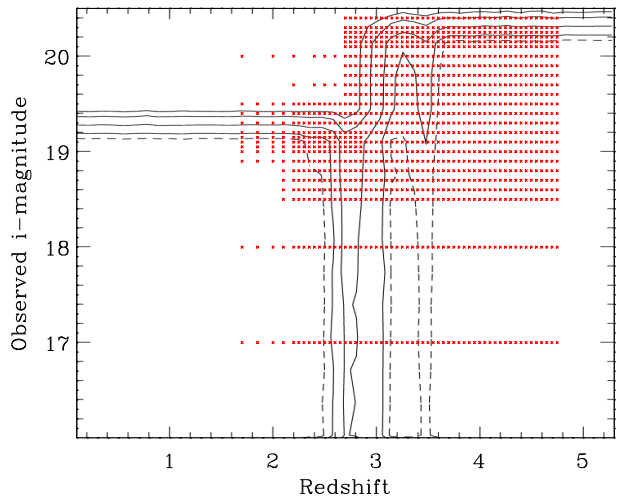}
  \caption{Redshift--magnitude combinations tested for completeness in
    the SDSS quasar target selection algorithm.  The contours show the
    completeness for non-BAL quasars measured by \citet{Richards02},
    marking the 10, 25, 50, 75 and 90 per cent levels.  The 90 per
    cent contour is marked with a dashed line.  
    Red crosses show the combinations tested in this work.}
  \label{fg:synspec.zi}
\end{figure}

Synthetic SDSS colours for the model quasars, over a range of
redshifts, were determined as described in \citet{Hewett06}. $ugriz$
magnitudes were then derived by setting the $i$-magnitude to a
specific value and applying the synthetic colours.  The magnitudes
were calculated according to the asinh magnitude system presented by
\citet{Lupton99}, to match the SDSS photometric data.  The range of
redshifts and $i$-magnitudes used is shown in
Fig.~\ref{fg:synspec.zi}; the values were chosen to concentrate on the
regions where completeness varies rapidly according to
\citet{Richards02}.

The errors in the photometric measurements can cause an object to
scatter into or out of the target selection region.  Typical errors
were derived from the `BEST' photometric values and errors in the SDSS
DR5 quasar catalogue \citep{Schneider07}. For each individual
photometric measurement all the measurements in that band with a
magnitude within 0.25 were extracted, along with their associated
errors.  The first and third quartile values of the extracted error
distribution were chosen to represent `good' and `poor' observing
conditions, respectively.  For each set of synthetic magnitudes ten
`good' and ten `poor' realisations were created by adding Gaussian
random noise with the appropriate 1-$\sigma$ errors.

The photometric errors of an object are used by the SDSS target
selection algorithm to assess the significance of an object's
deviation from the stellar locus in colour space. The relevant
magnitude errors required are those pertaining to the {\it scattered}
rather than {\it intrinsic} photometric properties.  Thus, the
magnitude errors were recalculated for each realisation using the
`TARGET' photometric data from \citet{Schneider07}, which better reflect the
information originally used for quasar target selection.
The resulting sets of magnitudes and errors were processed by the SDSS
quasar target selection algorithm, as radio-quiet stellar
(point-source) candidates, to determine the spectroscopic target status
of each model quasar realisation.  The results from each set of twenty
realisations were combined to estimate the completeness for that model
quasar.

There are a
number of criteria under which an object can be targeted; the {\tt
PRIMTARG} header item in each SDSS spectrum's datafile specifies whether an
object was targeted as a high-redshift quasar candidate (HIZ), a
low-redshift quasar candidate (LOWZ), a FIRST radio source or some
other candidate type.  Multiple targeting flags for a single source
are allowed.  The target selection algorithm is described in detail by
\citet{Richards02}.  Only the HIZ and LOWZ flags are used here: these
flags depend only in the photometric properties of each individual
object, and are the criteria under which the majority of SDSS quasars
were selected.

To reduce the CPU time and data transfer requirements of processing
the very large number of test points produced, after an initial set of
30\,000\,000 objects had been processed by the target selection
algorithm further objects were first compared to the initial set.  If
the two nearest neighbours in magnitude space to a new object had the
same {\tt PRIMTARG} values then this value was adopted for the new
object too.  If they disagreed, the new object was processed by the
normal target selection algorithm.  Tests on a random subsample of
objects suggested that comparison to the nearest neighbours gave the
correct result in 94 per cent of cases; given the large number of
objects included in each completeness datapoint this represents an
acceptable error rate.

\subsection{Completeness contours in redshift--magnitude space}

\label{ss:contours}

\begin{figure*}
  \includegraphics[width=168mm]{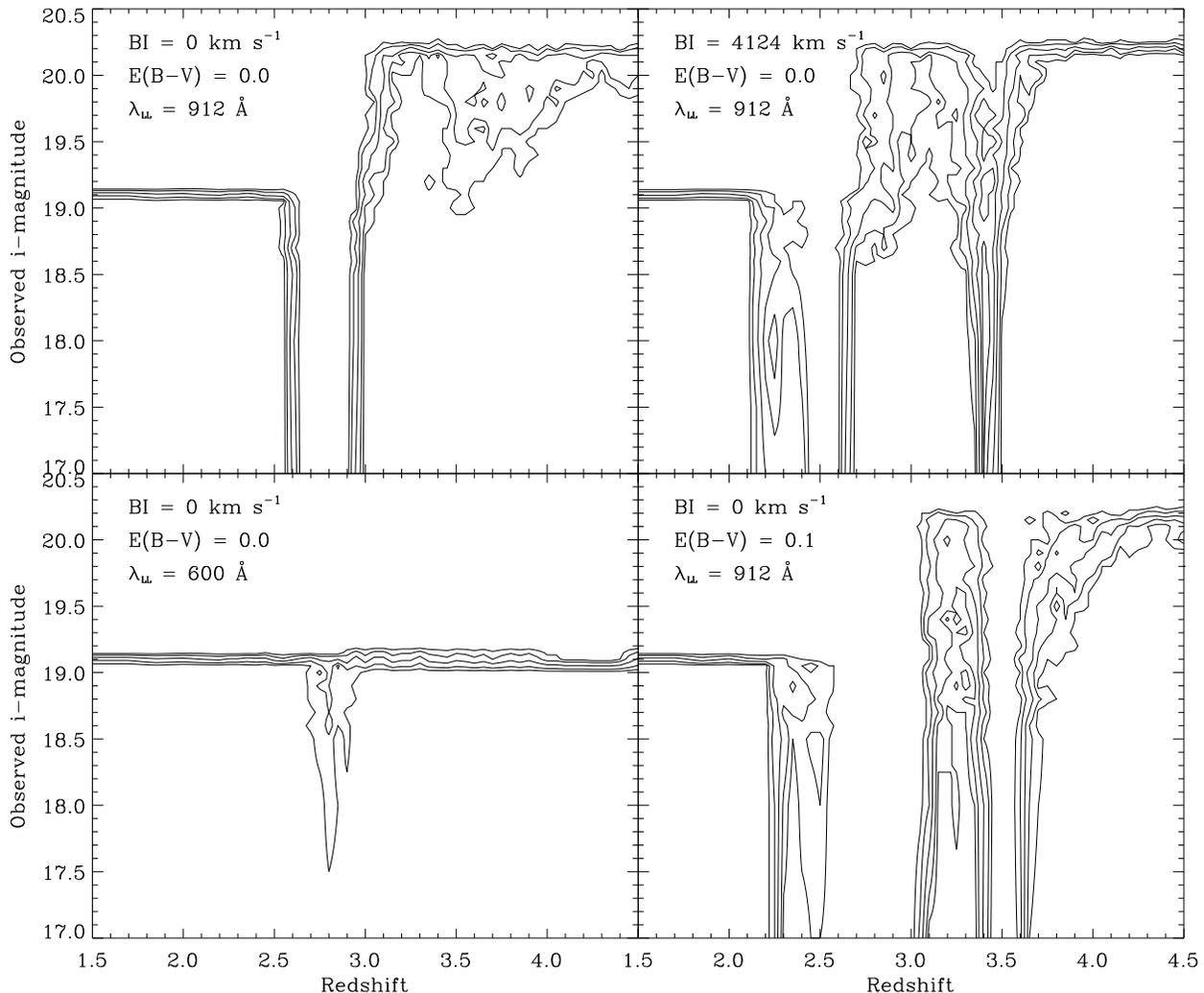}
  \caption{Example contour plots of the completeness derived for
    synthetic quasar spectra.  Contours correspond to completeness
    levels of 10, 25, 50, 75 and 90 per cent.  Top-left: non-BAL
    quasar, no dust-reddening, $\lambda_{\rm LL}=912$\,\AA.  Top-right:
    BALQSO with ${\rm BI} = 4124$\,\kms, no dust-reddening,
    $\lambda_{\rm LL}=912$\,\AA.  Bottom-left: non-BAL quasar, no
    dust-reddening, $\lambda_{\rm LL}=600$\,\AA.  Bottom-right: non-BAL
    quasar, $\ebv=0.1$, $\lambda_{\rm LL}=912$\,\AA.}
  \label{fg:comp}
\end{figure*}

Example completeness contour plots are shown in Fig.~\ref{fg:comp},
for the normal (not `blue') quasar SED with a range of input
properties.  The overall features for the dust-free quasars with
$\lambda_{\rm LL}=912$\,\AA\ (top-left panel for a non-BAL quasar,
top-right for a BALQSO) are largely as expected from the results of
\citet{Richards02} and \citet{Reichard03b}: the completeness is very
high for both non-BAL quasars and BALQSOs with $z < 2.1$ and $i <
19.1$, drops sharply for a narrow region close to $z$$\sim$2.6, and
rises again for higher redshifts.

Differences between the non-BAL and BALQSO completeness values can be
seen around $z$$\sim$2.6, where the model quasar colours move close to
or through the stellar locus.  The redder $u-g$ colours for BALQSOs,
caused by supression of the flux bluewards of 1050\,\AA, causes
BALQSOs to enter and then leave the stellar locus at lower redshifts
than non-BAL quasars.  A further difference is seen at $z$$\simeq$3.5,
where the BALQSO colours again come close to the stellar locus,
causing a sharp but brief drop in the completeness.  At other
redshifts, in particular $z$$<$2.1 and $z$$>$3.6, the completeness is
very similar for BALQSOs and non-BAL quasars.

Adjusting the position of the Lyman limit cut-off wavelength can have
a very strong effect on the derived completeness values, due to the
resulting change in the $u$-magnitude of the quasars.  The bottom-left
panel of Fig.~\ref{fg:comp} shows an extreme case in which a very low
$\lambda_{\rm LL}$=600\,\AA \ allows significant $u$-band flux at all
redshifts, keeping the quasar colours away from the stellar locus but
also preventing selection based on the HIZ targeting.

The contour plots in Fig.~\ref{fg:comp} are based on the {\em
  observed} $i$-band magnitude, so the bottom-right panel, with
$\ebv=0.1$, shows the effects of dust reddening but not dust
extinction.  However, there are still large differences between the
results for dust-free and dust-reddened quasars due just to the
induced change in colours.  \citet{Reichard03b} showed that, depending
on redshift, the effect of dust can be to move the quasar away from or
towards the stellar locus in colour space.  Such motion makes target
selection more or less likely at different redshifts.  The tendency
for BALQSOs to exhibit more dust reddening than non-BAL quasars
(e.g.\ \citealt{SF92,Reichard03b}) is thus important when considering
relative completeness values for the SDSS quasar selection.

Combining the different quasar parameters has effects on the
completeness maps that are not obvious from examining the parameters
individually.  The plots in Fig.~\ref{fg:comp} only illustrate the
types of effect that arise.  The impact of combinations of
dust-reddening, Lyman limit cut-off wavelength and BAL properties
must be considered in order to quantify the relative completeness.

\subsection{Distributions of quasar properties}

\label{ss:distributions}

In order to apply the completeness maps derived above, estimates must
be made of the distributions of the quasar properties on which they
are based.  Such distributions allow us to weight the results from
different input parameters accordingly. There are five properties for
which quasar population distributions are required: (i) the intrinsic
luminosity distribution of quasars, (ii) the intervening absorber
Lyman-limit wavelengths, (iii) the actual BAL trough properties, (iv)
the relative numbers of quasars with `normal' and `blue' SEDs and (v)
dust extinction and reddening, i.e.\ \ebv.  The first four of
these are covered here, while the \ebv\ distributions are described in
Secion~\ref{ss:ebv_dist}.

The intrinsic luminosity distribution was modelled as a power law,
$\Phi \left( \log \left( \lambda L_{1700} \right) \right) \propto
10^{-\alpha \log \left( \lambda L_{1700} \right)}$, with
$\alpha=2.2$.  The value of $\alpha$ was determined from a fit to the
observed distribution of quasars with $z<2.0$.  Within each redshift
bin the number density per unit redshift was assumed to be constant. 
As the completeness correction is applied to separate redshift bins
independently the change in number density between redshift bins does
not enter into the calculations.

The following results do not depend strongly on the details of the
luminosity distribution used.  The effect of using different values of
$\alpha$ is to change slightly the overall intrinsic BALQSO fraction
while making little or no difference to the shape of the fraction as a
function of redshift; the effects are discussed further in
Section~\ref{ss:target_corr}.  Tests using a redshift-dependent
luminosity distribution determined from the observed quasars in each
redshift bin produced very similar results.  The stability in the
determination of the intrinsic BALQSO fraction, $f_{\rm int}$
(Section~\ref{sc:intrinsic}), is in part because the final result
depends only on the {\em ratio} of the non-BAL quasar and BALQSO
completenesses, rather than their absolute values.

In order to weight the contributions from each of the Lyman limit
wavelengths an initial weight of 50 per cent was assigned to
$\lambda_{\rm LL}=912$\,\AA, to account for quasars where a Lyman
limit system exists within the host galaxy.  This fraction is
consistent with a visual inspection of SDSS quasar spectra.  To assign
the weights for the remaining half of the quasars, a set of 10$^6$
quasar sightlines were populated with Lyman limit systems and damped
Ly$\alpha$ absorbers using the method of \citet*{WHO94} and parameters
of \citet{Fan99}.  The completeness values derived for each of the
discrete $u$-magnitudes tested were interpolated to intervening
$u$-magnitudes, corresponding to different $\lambda_{\rm LL}$, and
weighted according to the fraction of the 10$^6$ quasar sightlines
whose highest-redshift absorber produced that value of $\lambda_{\rm
  LL}$.

The SDSS selection effects were calculated and applied separately for
non-BAL quasars and each of the 24 BALQSOs described in
Section~\ref{ss:synspec}.  The `normal' and `blue' quasar SEDs in the
population were assigned weights of 0.9 and 0.1, respectively.

\subsection{$\bmath{E(B-V)}$ distributions}

\label{ss:ebv_dist}

The \ebv\ probability density function for the non-BAL
quasars was taken from Maddox et al.\ (2010, in preparation). Maddox
et al.\ matched the SDSS quasar catalogue to UKIRT Infrared Deep Sky
Survey (UKIDSS; \citealt{UKIDSS}) $YJHK$ near-infrared
photometry. $E(B-V)$ estimates were then made for each quasar using
the observed SDSS $i$-band to near-infrared colours, relative to the
colours for the unreddened model quasar SED employed in this
paper. The method utilises quasar rest-frame wavelengths in the
optical, where the Milky Way, LMC and SMC extinction curves show very
similar behaviour. Maddox et al.\ (2010) take careful account of a
number of selection effects involved in combining the SDSS and UKIDSS
data but their resulting \ebv\ probability density functions
parametrize the {\em observed} distribution of extinction values for
quasars satisfying the SDSS LOWZ and HIZ quasar selection criteria.

The observed \ebv\ distribution is biased towards low values as, in
general, the presence of dust makes a quasar less likely to be
included in the SDSS spectroscopic survey. To recover the intrinsic
distribution of quasar properties, an estimate of the overall
completeness as a function of \ebv\ was made.  The observed
\ebv\ distribution was divided by the average completeness for a
population of quasars with the power-law luminosity distribution
described in Section~\ref{ss:distributions}, and the observed redshift
distribution, after applying the appropriate levels of reddening and
extinction.  The \ebv\ distribution was truncated at $\ebv = 0.2$ as
for larger values of $\ebv$ the completeness is too low to determine
the intrinsic distribution with sufficient accuracy.

\begin{figure}
  \includegraphics[width=84mm]{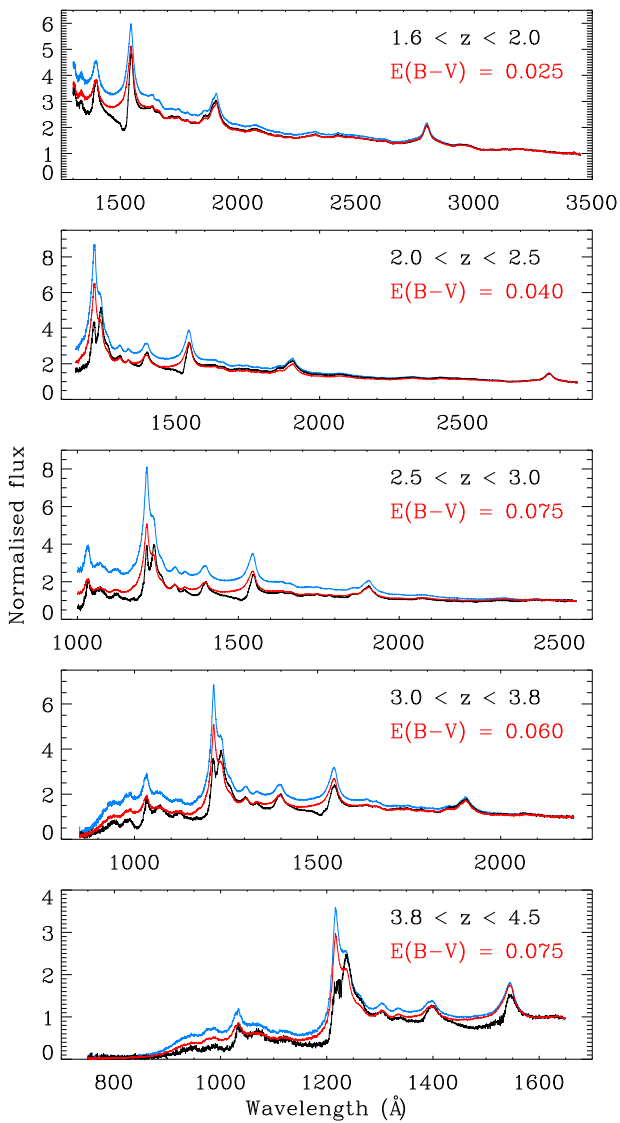}
  \caption{Composite spectra of BALQSOs and non-BAL quasars.  Each
    panel shows composites derived in a different redshift range, as
    labelled.  In each panel the black line shows the BALQSO composite,
    the blue line shows the non-BAL quasar composite, and the red line
    shows the same non-BAL composite after reddening by the
    \ebv\ level shown to match the shape of the BALQSO composite.  The
    \ebv\ measurements have uncertainties of $\pm$0.05 mag.  All
    spectra are normalised by the flux at the red end of the
    spectrum.  The numbers of BALQSO/non-BAL spectra used to generate
    each composite are, from low to high redshift, 1333/1204,
    900/1249, 497/1073, 538/1013 and 172/1226.  Non-BAL spectra were
    chosen at random from those available.}
  \label{fg:ebv_composites}
\end{figure}

In a flux limited sample, the mean observed \ebv\ for a population of
objects with an \ebv\ distribution independent of redshift, decreases
with redshift as the increasing level of extinction with redshift
experienced by the population removes a larger and larger fraction of
high \ebv\ objects.  By contrast, the mean observed \ebv\ for the
BALQSOs, derived from the SDSS $i$-band to near-infrared colours, {\em
  increases} strongly with redshift. Constructing composite non-BAL
and BALQSO spectra from the much larger sample of BALQSOs satisfying
the SDSS LOWZ and HIZ quasar selection criteria, as shown in
Fig.~\ref{fg:ebv_composites}, demonstrates the same effect: the
\ebv\ shown by the composite BALQSO spectra relative to the non-BAL
spectra increases significantly with redshift.

An observed variation in median \ebv\ could be caused by a genuine
trend with redshift, or by a redshift-dependent selection
effect. Indeed, the large \ebv=0.075 value found for the $z$=2.5--3.0
interval is explained by an increased probability of selecting quasars
with non-zero \ebv\ as the majority of unreddened quasars lie close to
the stellar locus in the $ugriz$ colour-space. However, the strong
increasing trend in median \ebv\ with redshift is present for redshift
ranges where the selection probabilities for both unreddened and
reddened quasars are extremely high. If dust content were strongly
correlated with BAL absorption strength this could produce an observed
correlation with redshift, as we are more sensitive to weak troughs at
low redshift, but an empirical determination of the relationship
between \ebv\ and BI for more than 300 BALQSOs in the redshift
interval 1.6$\le$$z$$<$2.6 shows no evidence for any such correlation.

Having ruled out any significant bias resulting from the quasar
selection or from a dependence of \ebv\ on BI, the most likely
explanation for the correlation between the observed median \ebv\ and
redshift is an increase in the intrinsic fraction of BALQSOs with
larger values of \ebv\ at higher redshifts.
Unfortunately, while the trend in the observed median \ebv\ for the
BALQSOs is clear we do not have sufficient information to determine
the form of the distribution of \ebv\ as a function of redshift.

We therefore adopt two related approaches to parametrizing the BALQSO
\ebv\ distribution.  In the first, a Gaussian distribution centred on
$\ebv$$=$$0.08$ with width, $\sigma$$=$$0.03$, truncated at
$\ebv$$=$$0.0$ and 0.2, was used, to provide a non-evolving reference.
The form of the distribution was chosen to produce an observed mean
\ebv=0.05 (over all redshifts 1.6$\le$$z$$<$4.5),
the approximate midpoint of the composite-derived values. For the
second approach, the Gaussian width was retained at 0.03,
independent of redshift, but a different central value was adopted for
each redshift bin such that the predicted mean observed \ebv\ matched
the linear trend with redshift derived from the composite quasar
spectra.  The central values of the Gaussian increase monotonically
from 0.023 at 1.65$\le$$z$$<$1.70 up to 0.099 at
4.0$\le$$z$$<$4.5. Hereafter, we refer to the two approaches as
`fixed' and `redshift-dependent', respectively.

A large number of alternative \ebv\ distributions were also tested to
determine the relationship between input \ebv\ distribution and the
final determination of the intrinsic BALQSO fraction.  As with
changing the luminosity distribution, modifying the form of the
\ebv\ distributions changes the overall BALQSO fraction but has little
effect on the shape of that fraction as a function of redshift as all
redshifts are affected in a similar manner; this point is discussed
further in Section~\ref{sc:discussion}.

\begin{figure}
  \includegraphics[width=84mm]{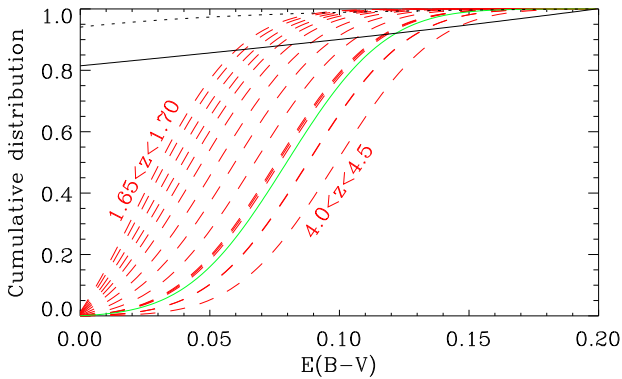}
  \caption{\ebv\ cumulative distribution functions.  Both the observed
    (dotted black line) and completeness-corrected (solid black line)
    CDFs for non-BAL quasars are shown.  For BALQSOs, the solid green
    line shows the fixed \ebv\ distribution, while dashed red lines
    show the redshift-dependent distributions.  The lowest and highest
    redshift bins are labelled; the CDFs vary monotonically with
    redshift in between.}
  \label{fg:ebv_dist}
\end{figure}

The \ebv\ cumulative distribution functions (CDFs) are shown in
Fig.~\ref{fg:ebv_dist}.  The CDFs for non-BAL quasars do not pass
through the origin because the probability distributions include a
delta function at $\ebv=0$ which, after correcting for completeness,
accounts for $\simeq$81 per cent of non-BAL quasars.

\subsection{Calculating overall completeness values}

\label{ss:completeness}

Overall completeness values were calculated in each
redshift--luminosity bin for non-BAL quasars and each of the 24
BALQSOs individually.  In short, quasars following the power-law
luminosity function were reddened according to the different
\ebv\ distributions described in Section~\ref{ss:distributions}.  The
completeness contours from Section~\ref{ss:contours} were then used to
determine the fraction of such quasars that would be targeted by the
SDSS, and the overall completeness was measured as the total number of
targeted quasars with a given observed luminosity divided by the total
input number of unreddened quasars with that luminosity.  This
approach enables the completeness calculation to explicitly include
reddened quasars with redshift--luminosity--\ebv\ combinations such
that their completeness is zero, which are not observed in the SDSS
catalogue.  Full details of the procedure are given below.  

First the quasar sample is restricted to those objects where reliable
detection probabilities are available.  The target selection testing
described above covers only the LOWZ and HIZ quasar target flags and
only those quasars
that were targeted under one or both of the LOWZ and HIZ criteria were
retained.  Early spectroscopic plates in the SDSS used slightly
different targeting criteria so in using all plates there is a small
bias against quasars that are not covered by the original targeting
criteria but would have been selected by the final algorithm.  This
bias can be removed by excluding plates that were observed before the
target selection algorithm was finalised.  Doing so makes no
significant difference to the results described below, suggesting the
systematic errors induced by the bias are very small, but the
statistical errors increase due to the reduced sample size.  As a
result, all spectroscopic plates are included in the sample.

\begin{figure}
  \includegraphics[width=84mm]{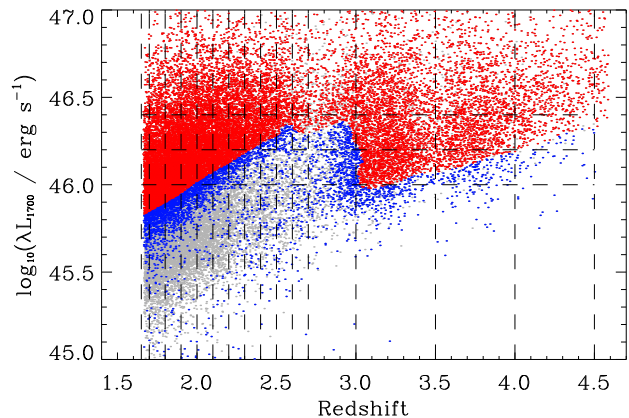}
  \caption{Redshift and luminosity values of SDSS quasars with
    complete coverage of the \civ\ region.  Grey points represent
    quasars that were not targeted under either the LOWZ or HIZ
    criteria; blue points lie in regions with completeness less than
    10 per cent; only the quasars represented by red points were used to
    calculate the intrinsic BALQSO fraction.  Dashed lines show the
    boundaries of the redshift--luminosity bins used.}
  \label{fg:zlcut}
\end{figure}

To reduce the impact of quasars with unusual spectral
properties, which were not covered by the target selection testing,
quasars with a redshift and apparent magnitude such that the completeness for
unreddened non-BAL quasars or BALQSOs (averaged over different trough
properties) was less than 10 per cent were excluded from the
calculations.  These quasars are by definition rare but their
inclusion would bias the calculation of the intrinsic BALQSO fraction
as these rare SEDs were not included in the calculation of the
completeness values.  Quasars with either the \civ\ Inc or
\civ\ BBP flags set were also rejected.  The redshift--luminosity
distribution of the 20\,078 retained quasars is shown in
Fig.~\ref{fg:zlcut}.

After restricting the input quasar sample the boundaries of the luminosity
ranges used to calculate $f_{\rm SDSS}$ were shifted to reflect the new
luminosity distribution; the new boundaries are also shown in
Fig.~\ref{fg:zlcut}.

The input \ebv\ distributions were used to derive the
redshift--luminosity distributions for quasars that are affected by
dust.  Intrinsic 1700-\AA\ luminosities were converted into observed
$i$-magnitudes for all \ebv\ values using extinction values and
$K$-corrections based on the appropriate synthetic non-BAL quasar and
BALQSO spectra.  From these magnitudes and the intrinsic luminosity
and \ebv\ distributions the number of quasars that would be targeted
by the SDSS was found from the completeness contours (derived above)
interpolated to the required redshift, observed magnitude and
\ebv\ values, and averaged over all redshifts within each bin.  The
completeness in each of the redshift--luminosity bins shown in
Fig.~\ref{fg:zlcut} was then calculated by dividing the number of
targeted quasars with an {\em observed} luminosity in that range by
the total number of quasars with an {\em intrinsic} luminosity in that
range.  The completeness was evaluated separately for non-BAL quasars
and each of the 24 BI/\md\ ranges for BALQSOs.

\section{Intrinsic BALQSO fraction}

\label{sc:intrinsic}

As detailed in Sections \ref{sc:p_det} and \ref{sc:selection}, to
derive the intrinsic BALQSO fraction, $f_{\rm int}$, from the observed
fraction, $f_{\rm obs}$, we must first correct for the incomplete
identification of BAL troughs within the SDSS catalogue, then correct
for the different levels of completeness for BALQSOs and non-BAL
quasars entering the SDSS quasar sample.  Over most regions of
parameter space these corrections are robust, but at particular
redshifts the probabilities quasars are selected become small,
systematic errors become large and the reliability of our analysis
decreases.  The regions where this occurs are discussed below.

\subsection{Correction for incomplete identification}

\label{ss:f_sdss}

An estimate of the true number of BALQSOs in the SDSS can be found by
splitting the observed BALQSOs into the redshift, luminosity and mean
depth bins used in Section~\ref{sc:p_det} and dividing the number in
each bin by the relevant detection probability.

\begin{figure}
  \includegraphics[width=84mm]{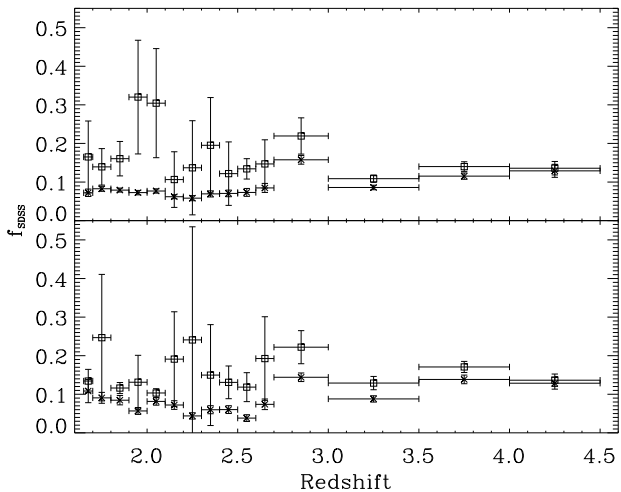}
  \caption{BALQSO fraction within the SDSS as a function of redshift.
    Upper panel: all luminosities.  Lower panel: extinction-corrected
    $\lambda L_{1700} \ge 10^{46.3}$\,erg\,s$^{-1}$.  Restricting the
    luminosities used minimizes the influence of the observed
    redshift--luminosity correlation.  In both panels the square
    symbols give the total fraction over all BAL troughs, while the
    crosses use $\md_{\rm cor} \ge 0.35$ only.}
  \label{fg:f_sdss_z}
\end{figure}

Using the redshift--luminosity bins shown in Fig.~\ref{fg:pdetbins}
and taking the total number of BALQSOs across all values of \md, the
resulting estimates of $f_{\rm SDSS}$ are shown as a function of
redshift as the square symbols in the upper panel of
Fig.~\ref{fg:f_sdss_z}.  The lower panel of Fig.~\ref{fg:f_sdss_z}
uses only quasars from the highest extinction-corrected luminosity bin
to ensure the use of the same luminosity interval over all redshifts.
The errors shown are 1-$\sigma$ errors calculated from bootstrap
realisations of the observed SDSS quasars and the quasars used to
calculate $p_{\rm det}$.

Summing quasars over all redshifts gives an overall fraction $f_{\rm
  SDSS} = 14.0 \pm 1.6$ per cent.  In this summation the two
redshift--luminosity bins in which the averaged $p_{\rm det}$ is less
than 0.05 were not used, as a small error in the value of $p_{\rm
  det}$ in these bins creates a much larger error in the resulting
calculation of $f_{\rm SDSS}$.

\begin{figure}
  \includegraphics[width=84mm]{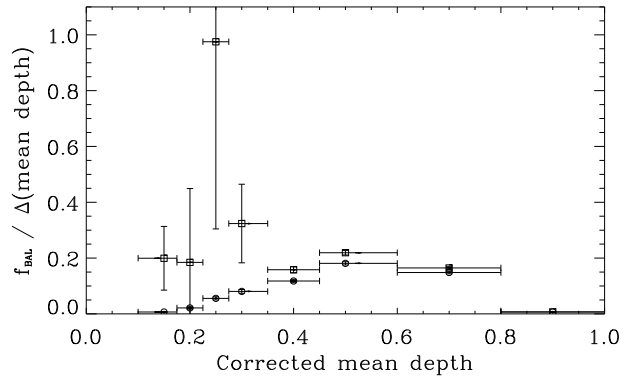}
  \caption{Observed (circles) and estimated SDSS (squares) BALQSO
    fractions as a function of corrected mean depth, binned in the
    $\md_{\rm cor}$ ranges shown by the horizontal error bars.  Symbols are
    plotted at the $\md_{\rm cor}$ values for which $p_{\rm det}$ was
    tested.  Note that many of the vertical error
    bars are smaller than the plotted symbols.}
  \label{fg:f_sdss_md}
\end{figure}

The statistical uncertainties in the determination of $f_{\rm SDSS}$
are very large in places; this is principally the result of large
uncertainties in the true number of BAL troughs with low $\md_{\rm
  cor}$.  For a number of datapoints the value plotted in
Fig.~\ref{fg:f_sdss_z} is dominated by just one or two BALQSOs that
lie in regions with very low $p_{\rm det}$.  The strong dependence of
the uncertainties on the mean depth can be seen in
Fig.~\ref{fg:f_sdss_md}, which shows $f_{\rm SDSS}$ as a function of
$\md_{\rm cor}$, averaged over all redshifts and luminosities.  The
results are binned according to the mean depths at which $p_{\rm det}$
was measured.  It is
clear that the small number of troughs observed with low \md\ in the
catalogue presented in Section~\ref{sc:results} is due to the small
detection probabilities, rather than an intrinsic lack of shallow
absorbers.

The crosses in Fig.~\ref{fg:f_sdss_z} show the results when only
BAL troughs with $\md_{\rm cor} \ge 0.35$ are included; the cut-off depth was
chosen to ensure a high detection probability over all redshifts and
luminosities, and hence a low uncertainty in the true number of
BALQSOs.  The resulting SDSS BALQSO fraction includes only a subset of
all BALQSOs, but the increased precision allows a better determination
of the redshift dependence.

\subsection{Correction for SDSS target selection}

\label{ss:target_corr}

The intrinsic numbers of non-BAL quasars and BALQSOs were calculated by
dividing the numbers of each estimated to be present in the SDSS by
the completeness values derived in Section~\ref{sc:selection}, using the
redshift--luminosity bins shown in Fig.~\ref{fg:zlcut}.  The
correction was applied separately for the fixed and redshift-dependent
\ebv\ distributions.

\begin{figure}
  \includegraphics[width=84mm]{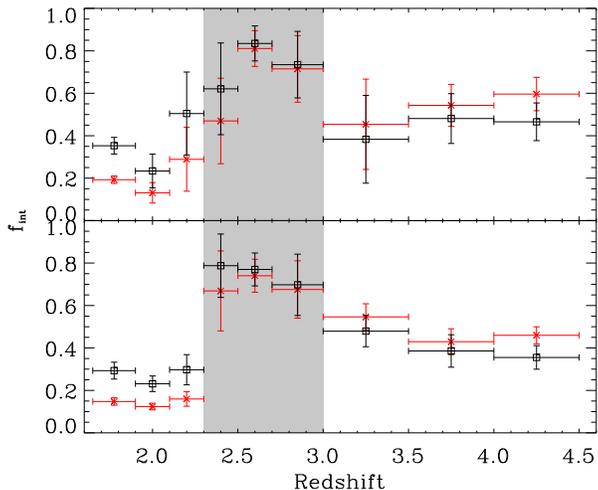}
  \caption{Intrinsic BALQSO fraction as a function of redshift,
    including troughs with $\md_{\rm cor} \ge 0.35$ only.  Top panel: all
    luminosities.  Bottom panel: extinction-corrected $\lambda
    L_{1700} \ge 10^{46.4}$\,erg\,s$^{-1}$ only.  Restricting the
    luminosities used minimizes the influence of the observed
    redshift--luminosity correlation.  In both panels black squares
    and red crosses use fixed and redshift-dependent
    \ebv\ distributions, respectively.  The shaded region marks the
    redshift range in which the systematic errors are predicted to be
    considerably larger than the statistical errors shown.}
  \label{fg:f_int_z}
\end{figure}

The intrinsic BALQSO fraction, $f_{\rm int}$, is shown as a function
of redshift in Fig.~\ref{fg:f_int_z}, in which the results from some
adjacent redshift bins have been combined to reduce the statistical
uncertainties.  The top panel uses quasars with all luminosities,
while the bottom panel uses only quasars with extinction-corrected
$\lambda L_{1700} \ge 10^{46.4}$\,erg\,s$^{-1}$ in order to minimize
the variation with redshift of the luminosity distribution.  In both
panels black squares represent results using the fixed
\ebv\ distribution, and red crosses use the redshift-dependent
distribution.  Only BAL troughs with $\md_{\rm cor} \ge 0.35$ were used, because
of the large uncertainties at lower values of $\md_{\rm cor}$ discussed in
Section~\ref{ss:f_sdss}.  As before, the errors are calculated from
bootstrap realisations of the observed SDSS quasars, the quasars used
to calculate $p_{\rm det}$, and also the sets of synthetic magnitudes
from which the target selection completeness contours were derived.

As well as the statistical errors shown the results will be affected
by systematic errors.  Details of tests carried out to determine the
level of systematic errors are given below but throughout most of the
redshift range covered these errors are relatively small.  However, in
the range 2.3$\le$$z$$<$3.0 the completeness for one or both of
non-BAL quasars and BALQSOs drops to very low levels, as previously
noted by \citet{Richards02} and \citet{Reichard03b}.  When the
completeness is low the systematic errors become far more significant,
for two reasons: firstly, a small absolute error in the completeness
measurement corresponds to a large relative error; and secondly, a
greater fraction of the observed objects will have unusual SEDs that
were not covered by the completeness testing described in
Section~\ref{sc:selection}.  As such, we caution that the systematic
errors for 2.3$\le$$z$$<$3.0 are likely to be considerably larger than
the statistical errors plotted in Fig.~\ref{fg:f_int_z}.  The affected
region is shaded in Fig.~\ref{fg:f_int_z}.  In other redshift regions
the completeness is high and hence the uncertainties are smaller, so
it is on these regions that we concentrate in the following discussion
(Section~\ref{sc:discussion}).

\begin{figure}
  \includegraphics[width=84mm]{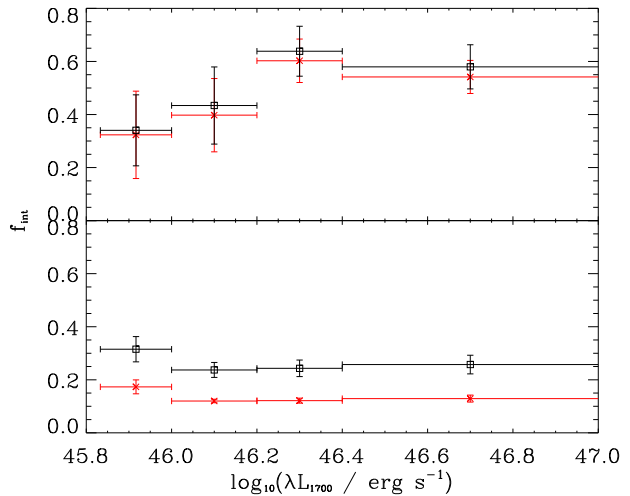}
  \caption{Intrinsic BALQSO fraction as a function of
    extinction-corrected luminosity, including troughs with $\md_{\rm
      cor} \ge 0.35$ only.  Top panel: all redshifts. Bottom panel:
    the results restricted to quasars with $z < 2.0$.  In both panels
    black squares and red crosses use fixed and redshift-dependent
    \ebv\ distributions, respectively.  Restricting the redshift range
    used minimizes the influence of the redshift--luminosity
    correlation, which is responsible for the strong trend evident in
    the top panel.}
  \label{fg:f_int_l}
\end{figure}

Fig.~\ref{fg:f_int_l} shows the results for $f_{\rm int}$ as a function of
extinction-corrected luminosity.  In this case the top panel uses data
from all redshifts while the bottom panel restricts the redshift range
to $z$$<$2.0.  The symbols and colours are the same as in
Fig.~\ref{fg:f_int_z}.

Taking the average over all luminosities and \md\ values, and all
redshifts $z$$<$2.3 or $z$$\ge$3.0, the intrinsic BALQSO fraction is
$f_{\rm int} = 38.8 \pm 2.2$ or $40.7 \pm 5.4$ per cent for the fixed
and redshift-dependent \ebv\ distributions, respectively.  In
calculating this fraction, redshift--luminosity bins were used only if
the overall completeness for both non-BAL quasars and BALQSOs, and the
averaged $p_{\rm det}$, were all greater than 0.05.
If the analysis is restricted to quasars that exhibit only strong BAL
troughs, with $\md_{\rm cor} \ge 0.35$, the intrinsic fraction falls
to $30.2 \pm 3.4$ or $25.6 \pm 2.7$ per cent for the two
\ebv\ schemes.

An important source of systematic error in the calculation of $f_{\rm
  int}$ is the determination of $p_{\rm det}$.  In particular, the
values of $p_{\rm det}$ used were averages over a range of BI values
at any particular $\md_{\rm in}$, so if the sample of BAL trough
profiles used was unrepresentative the determination of $p_{\rm det}$
would be biased.  Using each of the profiles in
Fig.~\ref{fg:fr.btdl.profiles} in isolation, rather than averaging
over all five, gives a range of values of $f_{\rm SDSS}$ and $f_{\rm
  int}$ with standard deviations of 4 and 8 per cent, respectively,
which gives an indication of the variation that could be induced by
making an extreme change to the distribution of BAL trough profiles.
Restricting the analysis to deep troughs ($\md_{\rm cor} \ge 0.35$)
decreases the possible variation in the results by a factor of 4, as
the typical corrections for incompleteness become smaller.

The input luminosity distribution used has an effect on the values of
$f_{\rm int}$ measured, principally because a steeper luminosity
distribution results in a larger fraction of reddened quasars dropping
below any luminosity limits for any given \ebv.  For the fixed
\ebv\ distribution, changing $\alpha$ in the luminosity distribution by 0.1
changes $f_{\rm int}$ by approximately 1 per cent at all redshifts,
with a steeper (shallower) luminosity distribution giving a larger
(smaller) $f_{\rm int}$.  For the redshift-dependent
\ebv\ distribution the effect is greatest at high redshift where the
typical \ebv\ values for BALQSOs are highest.  However, at all
redshifts the change in $f_{\rm int}$ is less than 1.7 per cent for a
change in $\alpha$ of 0.1, so an unfeasibly shallow luminosity
distribution would be required to remove the strong redshift evolution
seen in Fig.~\ref{fg:f_int_z}.

The results for the BALQSO fraction are not
strongly sensitive to the positions of the luminosity boundaries shown
in Figures \ref{fg:pdetbins} and \ref{fg:zlcut}: examining the range in
results produced using different positions suggested the systematic
error is no more than $\pm1.2$ per cent in $f_{\rm SDSS}$ and $\pm4.2$ per
cent in $f_{\rm int}$.  

Varying the relative fractions of the `regular' and `blue' model
quasars makes little difference to the calculated intrinsic BALQSO
fraction for $z<2.3$ and $z>3.0$, where the completeness for both
models is reasonably high. For the intermediate redshift interval,
where the SDSS quasar sample suffers from very significant
incompleteness, `blue' BALQSOs are significantly more likely to be
included than `regular' BALQSOs, so increasing the fraction of `blue'
quasars decreases the completeness correction for the BALQSOs and
hence decreases $f_{\rm int}$.

\section{Discussion}

\label{sc:discussion}

In the preceding Sections we have presented a detailed analysis of the
principal selection effects that affect the observed fraction of
high-ionisation BALQSOs (HiBALQSOs).  Even in the SDSS spectroscopic
samples, the number of LoBALQSOs at redshifts $z$$<$1.5 is relatively
small and we have not quantified the selection effects for such
objects.  However, the number of LoBALQSOs contained in our, and other
recent, catalogues suitable for a careful analysis of multiple
transitions from the same species (e.g.\ \citealt{Moe09}) is growing and a
better understanding of the frequency of occurrence and importance of
low-ionisation species in the BALQSO population should follow.

With the quantitative information in hand for the HiBALQSO population
we are able to measure the intrinsic BALQSO fraction and, more
importantly, its variation with redshift and luminosity.  Quantifying
such variation allows us to provide important constraints on BALQSO
models in a way not previously possible.

\subsection{Overall BALQSO fraction}

The observed \civ\ BALQSO fraction $f_{\rm obs} = 8.0 \pm 0.1$ per
cent derived here is lower than that from previous work, but the
resulting intrinsic fraction $f_{\rm int} = 38.8 \pm 2.2$ or $40.7 \pm
5.4$ per cent is
considerably larger than most other estimates.  The low observed
fraction is largely due to the decision not to smooth spectra before
calculating the BI, resulting in a more conservative definition of
BALQSOs.  This difference is accounted for in the correction for BAL
troughs that are present in the SDSS spectra but undetected when
continuum reconstructions are made. It is worth noting that the
effective `smoothing' of spectra prior to searches for BALQSOs has to
date been largely ad-hoc, resulting from either instrumental
limitations (i.e.\ resolution $R$=$\delta\lambda/\lambda$) or
small-scale ($\le$200\,\kms) filters (e.g.\ G09). An
alternative approach to that adopted in this paper would be to
optimise the detection of BALQSOs by applying a filtering scheme with
a scale $\simeq$2000\,\kms, i.e.\ equivalent to the minimum extent of BAL
troughs, thereby maximising the completeness of a BALQSO
catalogue. Part of the rationale for making available the
NMF-generated continua is indeed to allow such an approach to be
undertaken and relevant comparisons made between different `detection'
schemes for BAL troughs.

Recent work (\citealt*{Dai08}; \citealt{Maddox08}) has advocated the
existence of high BALQSO fractions based on samples selected at
near-infrared wavelengths.  However, the \citet{Dai08} type of
analysis is still dependent on the {\it observed} distribution of
quasars in the SDSS (optical) catalogues.  In the investigation
described here the optical-to-near-infrared properties of the
SDSS-selected quasars are first used to determine the {\it observed}
distributions of \ebv\ for non-BAL quasars, up to moderate reddenings
of \ebv=0.2 magnitudes.  For BALQSOs parametrizations of the intrinsic
\ebv\ distribution are chosen to reproduce the observed distribution,
accounting for the strong selection effects present.  Then, the
\ebv\ distributions are included in a self-consistent way to allow the
determination of the {\it intrinsic} fraction of BALQSOs (up to the
\ebv=0.2 mag limit).  The intrinsic fraction of BALQSOs derived here
is very similar to the {\it observed} $\simeq$40 per cent value from
\citet{Dai08}, based on a sample of SDSS quasars detected in 2MASS.
The observed fraction of BALQSOs will increase for flux-limited
samples defined at increasingly longer wavelengths. Determination of
the intrinsic fraction of BALQSOs from a sample involving flux limits
in two passbands, defined at substantially different epochs, involves
quantification of rather complex selection effects and we believe the
apparent agreement between the quoted BALQSO fractions to be somewhat
misleading.  \citet{Urrutia09} also propose a very high fraction of
BALQSOs based on a sample derived using FIRST, 2MASS and the SDSS.
Their objects possess \ebv\ values extending well beyond the \ebv=0.2
mag limit used here, and the number of quasars at $z$$>$1 is small.
The individual quasars are impressive examples of significantly
reddened objects but, again, a careful analysis of the selection
effects, including the strong bias towards the identification of red,
broad-line objects with strong H$\alpha$ emission present in the
$K$-band, for the redshift interval $z$$\simeq$2.1--2.6, is required.

In an analysis that is similar in concept to that described here,
\citet{Reichard03b}, whose results were also used in \citet{Knigge08},
the effect of larger \ebv\ values among BALQSOs was modelled by
taking the difference in colours between the SDSS EDR composite quasar
spectrum and a composite of HiBALQSOs.  This colour difference was
applied to sets of synthetic quasar colours to allow each object to be
processed by the SDSS quasar target selection algorithm as both a
BALQSO and non-BAL quasar, leading to a correction for
colour-dependent selection effects.  A separate correction was applied
to account for different levels of dust extinction.  The
\citet{Reichard03b} method produces an average completeness correction
based on quasars that were spectroscopically observed, but as such it
is biased against those quasars for which the SDSS completeness is
low, and it does not fully explore the parameter space of (reddened)
quasar properties.  Employing simulations of quasar SEDs for an
extended range of \ebv\ values, as described in Section
\ref{sc:selection}, and processing the reddening and extinction
corrections together, provides a more accurate determination of the
relative completeness levels.

The two different parametrizations of the \ebv\ distributions,
described in Section~\ref{ss:distributions}, produce consistent
results for the overall BALQSO fraction.  Such agreement is
unsurprising given the parametrizations were each chosen to reproduce
the typical \ebv\ values observed in the identified BALQSOs.  We note
that the overall BALQSO fraction is highly sensitive to the input
\ebv\ distribution: in general, $f_{\rm int}$ is correlated with the
mean input \ebv\ as a large fraction of high-\ebv\ objects would imply
a low completeness.  However, any input \ebv\ distribution used must be
consistent with the typical values observed, placing a strong
constraint on the possible range of distributions.

\subsection{Redshift dependence}

\label{ss:f_int_z}

Notwithstanding the quantification of the detection probabilities for
BAL troughs as a function of $\md$ (Fig.~\ref{fg:p_det}), the
combination of low $p_{\rm det}$ and small-number statistics mean that
the observed fraction of troughs with $\md_{\rm cor}$$<$0.35 is poorly
constrained as a function of any parameter of potential interest,
e.g.\ redshift.  Indeed, for $z$--$\lambda
L_{1700}$--\md\ combinations for which $p_{\rm det}$ drops low enough
there are no observed BALQSOs and hence we can at best provide an
upper limit to their numbers.  The problem can be seen in
Fig.~\ref{fg:f_sdss_z}, in which the $f_{\rm SDSS}$ values for all
BALQSOs, and only those with $\md_{\rm cor} \ge 0.35$, converge at
$z$$\ge$3.0, i.e.\ no additional shallow troughs are detected.

To investigate potential variation in the intrinsic fraction of
BALQSOs as a function of redshift and luminosity we therefore confine
ourselves to consideration of the BALQSO sample with $\md_{\rm
  cor}$$\ge$0.35.  When the BALQSO sample is limited in this way, as
shown in Fig.~\ref{fg:f_int_z}, $f_{\rm int}$ apparently peaks at $>$80 per
cent at $z$$\simeq$2.6.  However, the proximity of quasars to the
stellar locus in the SDSS $ugriz$-space at this redshift results in a
very low completeness for both BALQSOs and non-BAL quasars and we do
not believe the simulations of the type we have undertaken are
adequate for redshifts 2.3$\le$$z$$<$3.0. 
The larger median observed \ebv\ for the BALQSOs in the interval,
relative to adjacent redshifts (Section~\ref{ss:ebv_dist}), likely
indicates that further testing is required to fully explore the
sensitive linkage between quasar SED properties and the SDSS selection
at these redshifts.
Systematic uncertainties are thus considerably larger
than the statistical errors plotted in Fig.~\ref{fg:f_int_z} and in
the following discussion we focus on the results for $z$$<$2.3 and
$z$$\ge$3.0, where the completeness is considerably higher and hence
the uncertainties both smaller and well quantified.

Comparing the low ($z$$<$2.3) and high ($z$$\ge$3.0) redshift regions,
we see different patterns depending on the \ebv\ distributions used.
With a fixed \ebv\ distribution there is little overall difference
between the two regions, but the redshift-dependent distribution
results in a significantly higher $f_{\rm int}$ at high redshift.
This trend appears because a higher mean \ebv\ is necessary at high
redshift, to match the observed trend, and so a lower completeness for
high-$z$ BALQSOs is predicted.

An observed trend with redshift can in general be produced by an
intrinsic trend with luminosity, although the multiple criteria under
which SDSS quasars were selected to some extent reduce the strong
redshift--luminosity correlation present in most flux-limited samples.
The results described above do not change greatly when we restrict our
sample to the most luminous quasars (bottom panel of
Fig.~\ref{fg:f_int_z}), but the trend of higher $f_{\rm int}$ at
higher redshift for the redshift-dependent \ebv\ distribution is
strengthened, and a similar but weaker trend can also be seen for the
fixed \ebv\ distribution.  With the luminosity threshold in place
there is very little redshift evolution of the luminosity distribution
of the quasars, implying that the observed trends represent a true
redshift evolution of the BALQSO fraction.

A different parametrization of the \ebv\ distribution as a function of
redshift would in general predict a different trend in $f_{\rm int}$.
However we note that, even when using the fixed \ebv\ distribution,
which overpredicts the mean \ebv\ at low redshift and underpredicts it
at high redshift, $f_{\rm int}$ at $z$$\ge$3.0 is greater than that at
$z$$<$2.3 by a factor 1.6$\pm$0.2.
To remove this trend would require a decreased
completeness at low redshift, or an increased completeness at high
redshift, either of which would exacerbate the disagreement between
the predicted and observed mean \ebv.  When using a redshift-dependent
\ebv\ distribution that better predicts the SEDs of composite BALQSO
spectra the redshift trend becomes considerably stronger, with a
factor 3.5$\pm$0.4 difference between the high- and low-redshift
intrinsic fractions.

Other than the \ebv\ distributions, the largest source of systematic
error we have identified is in
the choice of synthetic BAL troughs used to quantify the detection
probability, $p_{\rm det}$.  However, any variation in $p_{\rm det}$
due to a different range of synthetic troughs would affect all
redshifts in approximately the same way, making little difference to
the form of the observed trends.  Other sources of systematic error,
such as the luminosity distribution used or the positioning of the
redshift--luminosity bins, were found to have minimal effect on the
results.  Further sources of systematic error, of which we are aware,
would, to at least first order, also affect all redshifts in a similar
manner.  As such we expect the trends to be robust against known
sources of systematic error and Fig.~\ref{fg:f_int_z}, showing the
large, factor $\simeq$3.5, change in the fraction of BALQSOs as a function
of redshift, is the main scientific
result of our investigation.

A further implication of the high BALQSO fraction at high redshift is
in the number density of high-redshift quasars.  The relatively low
completeness for $z$$\ge$3.0 BALQSOs implies that the true quasar
number density at such redshifts is higher than observed by $\simeq$50
per cent. At face value the increase in space density is relatively
modest but if the results of \citet{Glikman10} concerning the
steepness of the faint end of the quasar luminosity function at
$z$$\sim$4 are confirmed then our result may have some implications in
the context of the ability of the quasar population to maintain the
ionisation of the inter-galactic medium at $z$$\le$5.

\subsection{Luminosity dependence}

There is a dynamic range of approximately one decade in luminosity at
fixed redshift in our sample (Fig.~\ref{fg:zlcut}). Somewhat unusually
for an optically defined quasar sample, the two different selection
algorithms employed in the SDSS (to target low- and high-redshift
quasars), with their different faint magnitude limits, result in a
relatively weak correlation between redshift and median luminosity. It
is thus viable to determine whether luminosity contributes to the very
strong redshift-dependent trends in BALQSO fraction discussed above.

The observed BALQSO fractions for \siiv, \civ\ and \aliii\ are
strongly dependent on luminosity, with a much higher fraction at high
values of observed $\lambda L_{1700}$.  This is in part because the
spectra of low-luminosity quasars will in general have lower S/N, and
a BAL trough is less likely to be identified in a low S/N spectrum.
However, the S/N dependence is to some extent balanced by the tendency
for BALQSOs to possess larger \ebv\ values (due to dust) than non-BAL
quasars, reducing the observed BALQSO luminosities.  After correcting
for the competing selection effects for \civ\ BALQSOs the intrinsic
fraction still shows a positive correlation with extinction-corrected
luminosity, but the apparent trend is the result of the redshift
dependence discussed in Section~\ref{ss:f_int_z}.  When the analysis
is restricted to quasars with $z$$<$2.0 and BAL troughs with
$\md_{\rm cor}$$\ge$0.35, $f_{\rm int}$ shows no dependence on
luminosity. However, the limited dynamic range in luminosity means
that the constraint on any luminosity-dependent behaviour is weak.
Fitting linear regression lines to the data in the lower panel of
Fig.~\ref{fg:f_int_l} gives slopes of 
$-2.1$$\pm$6.4 per cent dex$^{-1}$ (fixed \ebv\ distribution) and
0.1$\pm$2.3 per cent dex$^{-1}$ (redshift-dependent
\ebv\ distribution).  For the latter distribution the uncertainties
correspond to 3-$\sigma$ limits of $-6.9$ and 7.0 per cent dex$^{-1}$.

The observed fraction of \mgii\ BALQSOs is highest at the lowest
observed luminosities, in contrast to the other ions, due largely to a
population of high-\ebv\ objects.  The most extreme objects of this
type are often undetected at redshifts $z$$>$1.8 due to the
catastrophic reduction in the observed $i$-band flux once the absorbed
part of the quasar SED, shortward of \mgii, falls in the SDSS $i$
band. Dust-induced extinction is also likely to make the population of
\mgii\ BALQSOs even fainter as the quasar redshift
increases. Observational strategies of the type described by
\citet{Urrutia09} should prove far more effective in quantifying the
fraction of such extreme BALQSOs.

\subsection{Comparison to models}

One of the two primary classes of model for BALQSOs \citep{Weymann91}
involves the presence of broad absorption line clouds in all quasars
but, due to incomplete solid angle coverage, quasars are only observed
as BALQSOs when viewed along certain sightlines.  Hence, broad
absorption troughs are observed in some quasar spectra but not others,
and the fraction in which they are observed can be directly related to
the fractional solid angle coverage of the broad absorption regions.
\citet{Ganguly08} summarise the body of evidence that the BALQSO
fraction, and more generally the outflow fraction, is largely
independent of the properties of the quasar.  If incomplete solid
angle coverage were the only determinant of the intrinsic fraction of
BALQSOs it would be expected that the fraction should not vary as a
function of redshift: the geometry of the BAL region is predicted to
be constant with respect to time.

The results of Section~\ref{sc:intrinsic} provide evidence against
this simple model.  In particular, the intrinsic BALQSO fraction is
found to be significantly greater at high redshift ($z \ge 3.0$) than
at low redshift ($1.65 \le z < 2.3$).  Such a change requires a model
in which $f_{\rm int}$ depends on one or more parameters that are
themselves varying functions of redshift.  One possibility is that the
coverage of the broad absorption line regions varies during the life
of a quasar.  The probability of viewing a BAL region in any
particular quasar may still be a function of the solid angle coverage
of the BAL clouds, but that coverage would itself be a function of the
age of the quasar. \citet{Farrah07} and \citet{Urrutia09} have argued
recently that at least the most extreme examples of the BAL
phenomenon, the rare FeLoBAL quasars, are indeed the manifestation of
an evolutionary scheme in which the BALQSOs represent the final phase
in the emergence of a naked quasar from an earlier dust- and
gas-enshrouded fuelling phase.  The observed variation in \ebv\ with
redshift would also be consistent with such a scheme.

An acknowledged difficulty with the evolutionary class of models is
the lack of evidence for differences in the mid-infrared fluxes of
BALQSOs and non-BAL quasars \citep{Gallagher07}, expected to arise
from the presence of an obscuring `cocoon' at early times in the
evolution of the objects.  However, the evolutionary scenario has
attracted support from recent studies at other wavelengths (e.g.\
\citealt{MontenegroMontes09}), although observations are still at an
exploratory stage in many cases \citep{Priddey07}.  Our results relate
to the much more common HiBALQSOs but the large, factor of 3.5,
decrease in the BAL fraction from the highest redshifts,
$z$$\simeq$4.5, to redshift, $z$$\sim$2.6, where the observed space
density of the most luminous quasars peaks, coincides with the period
when the individual black hole growth within quasars was at its most
rapid.

Alternative models that allow for cosmic evolution of the BALQSO
fraction include identifying BAL regions with radiation-driven disc
winds (e.g.\ \citealt*{Proga00}; \citealt{Risaliti09}).  Such winds
are a class of outflow that can be generated by quasar accretion discs
under a variety of physical conditions.  As the solid angle coverage
of the winds, and indeed the possibility of their existence, is a
function of physical parameters such as the mass and Eddington ratio
of the quasar \citep{Proga04,Risaliti09}, and the typical values of
these parameters vary with redshift (e.g.\ \citealt*{Hopkins07};
\citealt{SE10a,SE10b}), models of this type would generate a
redshift-dependent BAL fraction without requiring -- but also without
contradicting -- evolutionary BALQSO models.

Although we have provided a very brief summary of some of the
considerations that relate to the main classes of models for BALQSOs
the primary purpose of this paper is to present the first quantitative
determination of the BAL fraction of luminous quasars as a function of
redshift and luminosity. At face value, the very strong systematic
changes in the BAL fraction as a function of redshift present a
significant challenge for current models.

\section{Conclusions}

\label{sc:conclusions}

We have applied non-negative matrix factorisation to the
reconstruction of SDSS quasar spectra, with particular reference to
BALQSOs, and presented the resulting measurements of BAL properties.
The observed \civ\ BAL fraction was corrected for incomplete
identification of BAL troughs by the NMF routine, and for differential
SDSS spectroscopic target selection for BALQSOs and non-BAL quasars.

The principal results from this analysis are that:
\begin{enumerate}
  \item A total of 811 \siiv, 3296 \civ, 214 \aliii\ and 215
    \mgii\ BALQSOs are detected, corresponding to observed BALQSO
    fractions of 3.4 $\pm$ 0.1, 8.0 $\pm$ 0.1, 0.38 $\pm$ 0.03 and
    0.29 $\pm$ 0.02 per cent, respectively.
  \item The probability of a BAL trough being detected by the NMF
    procedure is strongly dependent on the S/N of the spectrum and the
    mean depth of the trough.  The detection probability is often
    very low for $\md_{\rm cor}$$<$0.35, resulting in an observed deficit of
    shallow BAL troughs.
  \item After correcting for incomplete identification of BAL troughs,
    the estimated \civ\ BALQSO fraction within the SDSS spectroscopic
    survey is 14.0 $\pm$ 1.6 per cent.
  \item After correcting for differential SDSS target selection of BALQSOs
    and non-BAL quasars the estimated intrinsic \civ\ BALQSO fraction is
    40.7$\pm$5.4 per cent when using a redshift-dependent
    \ebv\ distribution.
  \item The intrinsic BALQSO fraction decreases by a factor of 3.5$\pm$0.4 
    between the redshift intervals 3.0$\le$$z$$<$4.5
    and 1.65$\le$$z$$<$2.3, implying that the orientation of a
    sightline with respect to the quasar and its torus alone is insufficient
    to determine the presence or otherwise of a BAL trough.
  \item The intrinsic BALQSO fraction shows no significant variation with the
    luminosity of the quasars in the sample, within the restricted
    luminosity range in which the comparison can be made.
\end{enumerate}

The NMF reconstructions in the region around the \civ\ emission and
absorption lines, of each of the 48\,146 quasar spectra in the sample
with coverage of the \civ\ region, will be made available through the
SDSS value added catalogues as this paper is published.

\section*{Acknowledgments}

We thank the referee, Patrick Hall, for a number of suggestions that
greatly improved the quality of this work.  We thank Vivienne Wild for
her contribution to improving the SDSS DR6 spectra via the
sky-residual subtraction scheme, and Michael Strauss for processing a
number of objects through the SDSS target selection algorithm.  JTA
acknowledges the award of an STFC Ph.D.\ studentship.  PCH
acknowledges support from the STFC-funded Galaxy Formation and
Evolution programme at the Institute of Astronomy.  GTR was supported
in part by an Alfred P.\ Sloan Research Fellowship.

Funding for the SDSS and SDSS-II has been provided by the Alfred
P. Sloan Foundation, the Participating Institutions, the National
Science Foundation, the U.S. Department of Energy, the National
Aeronautics and Space Administration, the Japanese Monbukagakusho, the
Max Planck Society, and the Higher Education Funding Council for
England. The SDSS Web Site is http://www.sdss.org/.

The SDSS is managed by the Astrophysical Research Consortium for the
Participating Institutions. The Participating Institutions are the
American Museum of Natural History, Astrophysical Institute Potsdam,
University of Basel, University of Cambridge, Case Western Reserve
University, University of Chicago, Drexel University, Fermilab, the
Institute for Advanced Study, the Japan Participation Group, Johns
Hopkins University, the Joint Institute for Nuclear Astrophysics, the
Kavli Institute for Particle Astrophysics and Cosmology, the Korean
Scientist Group, the Chinese Academy of Sciences (LAMOST), Los Alamos
National Laboratory, the Max-Planck-Institute for Astronomy (MPIA),
the Max-Planck-Institute for Astrophysics (MPA), New Mexico State
University, Ohio State University, University of Pittsburgh,
University of Portsmouth, Princeton University, the United States
Naval Observatory, and the University of Washington.

\bibliography{bibtrunc}{}
\bibliographystyle{mn2e}

\begin{table*}
  \caption{Properties of the NMF reconstructions and the resulting broad absorption properties.}
  \label{tb:results}
  \begin{tabular}{rrrrrrrrrr}
    \hline
    \multicolumn{1}{c}{SDSS object name } & \multicolumn{1}{c}{RA (J2000) } & \multicolumn{1}{c}{Dec. (J2000) } & \multicolumn{1}{c}{MJD } & \multicolumn{1}{c}{Plate } & \multicolumn{1}{c}{Fiber } & \multicolumn{1}{c}{$z$ } & \multicolumn{1}{c}{$i$ } & \multicolumn{1}{c}{$SN_{1700}$ } & \multicolumn{1}{c}{$\log(F_{1700})$ } \\
    \multicolumn{1}{c}{ } & \multicolumn{1}{c}{ deg } & \multicolumn{1}{c}{ deg } & \multicolumn{1}{c}{ } & \multicolumn{1}{c}{ } & \multicolumn{1}{c}{ } & \multicolumn{1}{c}{ } & \multicolumn{1}{c}{ mag } & \multicolumn{1}{c}{ } & \multicolumn{1}{c}{ [erg\,cm$^{-2}$\,s$^{-1}$\,\AA$^{-1}$] } \\
    \hline
    000006.53$+$003055.2 &   0.027231 &   0.515332 & 52203 &  685 & 467 & 1.8240 & 20.041 &   4.288 & -16.019 \\
    000008.13$+$001634.6 &   0.033946 &   0.276292 & 52203 &  685 & 470 & 1.8366 & 19.420 &   5.010 & -15.952 \\
    000009.26$+$151754.5 &   0.038609 &  15.298489 & 52251 &  751 & 354 & 1.1971 & 19.058 &   0.000 &   0.000 \\
    000009.38$+$135618.4 &   0.039099 &  13.938458 & 52235 &  750 &  82 & 2.2400 & 18.172 &  13.950 & -15.287 \\
    000009.42$-$102751.9 &   0.039264 & -10.464410 & 52143 &  650 & 199 & 1.8520 & 18.700 &   9.899 & -15.530 \\
    \hline
  \end{tabular}
\end{table*}

\begin{table*}
  \contcaption{}
  \begin{tabular}{rrrrrrrrrrrr}
    \hline
    \multicolumn{1}{c}{$\log(\lambda L_{1700})$ } & \multicolumn{1}{c}{HIZ } & \multicolumn{1}{c}{LOWZ } & \multicolumn{1}{c}{Primary } & \multicolumn{1}{c}{GenComp } & \multicolumn{1}{c}{$N_{\rm comp}$ } & \multicolumn{1}{c}{RedComp } & \multicolumn{1}{c}{$\chi_\nu^2$ } & \multicolumn{1}{c}{$N_{\rm pix}$ } & \multicolumn{1}{c}{Slope } & \multicolumn{1}{c}{SlCor } & \multicolumn{1}{c}{DipMask } \\
    \multicolumn{1}{c}{ [erg\,s$^{-1}$] } & \multicolumn{1}{c}{ } & \multicolumn{1}{c}{ } & \multicolumn{1}{c}{ } & \multicolumn{1}{c}{ } & \multicolumn{1}{c}{ } & \multicolumn{1}{c}{ } & \multicolumn{1}{c}{ } & \multicolumn{1}{c}{ } & \multicolumn{1}{c}{ } & \multicolumn{1}{c}{ } & \multicolumn{1}{c}{ } \\
    \hline
    45.574 & 0 & 0 & 1 & 0 & 12 & 0 &   0.869 & 3641 & -0.277 & 0 & 0 \\
    45.648 & 0 & 0 & 1 & 0 & 12 & 0 &   0.951 & 3589 &  0.576 & 0 & 0 \\
     0.000 & 0 & 1 & 1 & 0 &  9 & 0 &   1.175 & 3502 &  0.746 & 0 & 0 \\
    46.525 & 0 & 1 & 1 & 0 & 11 & 0 &   1.358 & 3468 &  0.109 & 0 & 0 \\
    46.080 & 0 & 1 & 1 & 0 & 12 & 0 &   0.973 & 3534 & -0.142 & 0 & 0 \\
    \hline
  \end{tabular}
\end{table*}

\begin{table*}
  \contcaption{}
  \begin{tabular}{rrrrrrrrrr}
    \hline
    \multicolumn{1}{c}{ManMask } & \multicolumn{1}{c}{\siiv\ BI } & \multicolumn{1}{c}{\siiv\ \md } & \multicolumn{1}{c}{\siiv\ $v_{\rm min}$ } & \multicolumn{1}{c}{\siiv\ $v_{\rm max}$ } & \multicolumn{1}{c}{\siiv\ $v_{\rm cov,min}$ } & \multicolumn{1}{c}{\siiv\ $v_{\rm cov,max}$ } & \multicolumn{1}{c}{\siiv\ $N_{\rm BR}$ } & \multicolumn{1}{c}{\siiv\ $N_{\rm BT}$ } & \multicolumn{1}{c}{\siiv\ $N_{\rm SR}$ } \\
    \multicolumn{1}{c}{ } & \multicolumn{1}{c}{ \kms } & \multicolumn{1}{c}{ } & \multicolumn{1}{c}{ \kms } & \multicolumn{1}{c}{ \kms } & \multicolumn{1}{c}{ \kms } & \multicolumn{1}{c}{ \kms } & \multicolumn{1}{c}{ } & \multicolumn{1}{c}{ } & \multicolumn{1}{c}{ } \\
    \hline
    0 &     0.0 & 0.000 &      0 &      0 &  -3000 &  -9108 &   0 &   0 &  0 \\
    0 &     0.0 & 0.000 &      0 &      0 &  -3000 &  -9108 &   0 &   0 &  0 \\
    0 &     0.0 & 0.000 &      0 &      0 &      0 &      0 &   0 &   0 &  0 \\
    0 &     0.0 & 0.000 &      0 &      0 &  -3000 & -25000 &   0 &   0 &  0 \\
    0 &     0.0 & 0.000 &      0 &      0 &  -3000 &  -9108 &   0 &   0 &  0 \\
    \hline
  \end{tabular}
\end{table*}

\begin{table*}
  \contcaption{}
  \begin{tabular}{rrrrrrrrrr}
    \hline
    \multicolumn{1}{c}{\siiv\ $N_{\rm ST}$ } & \multicolumn{1}{c}{\siiv\ Inc } & \multicolumn{1}{c}{\siiv\ CBP } & \multicolumn{1}{c}{\siiv\ CPF } & \multicolumn{1}{c}{\siiv\ BBP } & \multicolumn{1}{c}{\civ\ BI } & \multicolumn{1}{c}{\civ\ \md } & \multicolumn{1}{c}{\civ\ $v_{\rm min}$ } & \multicolumn{1}{c}{\civ\ $v_{\rm max}$ } & \multicolumn{1}{c}{\civ\ $v_{\rm cov,min}$ } \\
    \multicolumn{1}{c}{ } & \multicolumn{1}{c}{ } & \multicolumn{1}{c}{ } & \multicolumn{1}{c}{ } & \multicolumn{1}{c}{ } & \multicolumn{1}{c}{ \kms } & \multicolumn{1}{c}{ } & \multicolumn{1}{c}{ \kms } & \multicolumn{1}{c}{ \kms } & \multicolumn{1}{c}{ \kms } \\
    \hline
     0 & 1 & 0 & 0 & 0 &     0.0 & 0.000 &      0 &      0 &  -3000 \\
     0 & 1 & 0 & 0 & 0 &     0.0 & 0.000 &      0 &      0 &  -3000 \\
     0 & 1 & 0 & 0 & 0 &     0.0 & 0.000 &      0 &      0 &      0 \\
     0 & 0 & 0 & 0 & 0 &     0.0 & 0.000 &      0 &      0 &  -3000 \\
     0 & 1 & 0 & 0 & 0 &     0.0 & 0.000 &      0 &      0 &  -3000 \\
    \hline
  \end{tabular}
\end{table*}

\begin{table*}
  \contcaption{}
  \begin{tabular}{rrrrrrrrrr}
    \hline
    \multicolumn{1}{c}{\civ\ $v_{\rm cov,max}$ } & \multicolumn{1}{c}{\civ\ $N_{\rm BR}$ } & \multicolumn{1}{c}{\civ\ $N_{\rm BT}$ } & \multicolumn{1}{c}{\civ\ $N_{\rm SR}$ } & \multicolumn{1}{c}{\civ\ $N_{\rm ST}$ } & \multicolumn{1}{c}{\civ\ Inc } & \multicolumn{1}{c}{\civ\ CBP } & \multicolumn{1}{c}{\civ\ CPF } & \multicolumn{1}{c}{\civ\ BBP } & \multicolumn{1}{c}{\aliii\ BI } \\
    \multicolumn{1}{c}{ \kms } & \multicolumn{1}{c}{ } & \multicolumn{1}{c}{ } & \multicolumn{1}{c}{ } & \multicolumn{1}{c}{ } & \multicolumn{1}{c}{ } & \multicolumn{1}{c}{ } & \multicolumn{1}{c}{ } & \multicolumn{1}{c}{ } & \multicolumn{1}{c}{ \kms } \\
    \hline
    -25000 &   0 &   0 &  0 &  0 & 0 & 0 & 0 & 0 &     0.0 \\
    -25000 &   0 &   0 &  0 &  0 & 0 & 0 & 0 & 0 &     0.0 \\
         0 &   0 &   0 &  0 &  0 & 1 & 0 & 0 & 0 &     0.0 \\
    -25000 &   0 &   0 &  0 &  0 & 0 & 0 & 0 & 0 &     0.0 \\
    -25000 &   0 &   0 &  0 &  0 & 0 & 0 & 0 & 0 &     0.0 \\
    \hline
  \end{tabular}
\end{table*}

\begin{table*}
  \contcaption{}
  \begin{tabular}{rrrrrrrrr}
    \hline
    \multicolumn{1}{c}{\aliii\ \md } & \multicolumn{1}{c}{\aliii\ $v_{\rm min}$ } & \multicolumn{1}{c}{\aliii\ $v_{\rm max}$ } & \multicolumn{1}{c}{\aliii\ $v_{\rm cov,min}$ } & \multicolumn{1}{c}{\aliii\ $v_{\rm cov,max}$ } & \multicolumn{1}{c}{\aliii\ $N_{\rm BR}$ } & \multicolumn{1}{c}{\aliii\ $N_{\rm BT}$ } & \multicolumn{1}{c}{\aliii\ $N_{\rm SR}$ } & \multicolumn{1}{c}{\aliii\ $N_{\rm ST}$ } \\
    \multicolumn{1}{c}{ } & \multicolumn{1}{c}{ \kms } & \multicolumn{1}{c}{ \kms } & \multicolumn{1}{c}{ \kms } & \multicolumn{1}{c}{ \kms } & \multicolumn{1}{c}{ } & \multicolumn{1}{c}{ } & \multicolumn{1}{c}{ } & \multicolumn{1}{c}{ } \\
    \hline
    0.000 &      0 &      0 &  -3000 & -25000 &   0 &   0 &  0 &  0 \\
    0.000 &      0 &      0 &  -3000 & -25000 &   0 &   0 &  0 &  0 \\
    0.000 &      0 &      0 &  -3000 &  -8126 &   0 &   0 &  0 &  0 \\
    0.000 &      0 &      0 &  -3000 & -25000 &   0 &   0 &  9 &  0 \\
    0.000 &      0 &      0 &  -3000 & -25000 &   0 &   0 &  0 &  0 \\
    \hline
  \end{tabular}
\end{table*}

\begin{table*}
  \contcaption{}
  \begin{tabular}{rrrrrrrrrr}
    \hline
    \multicolumn{1}{c}{\aliii\ Inc } & \multicolumn{1}{c}{\aliii\ CBP } & \multicolumn{1}{c}{\aliii\ CPF } & \multicolumn{1}{c}{\aliii\ BBP } & \multicolumn{1}{c}{\mgii\ BI } & \multicolumn{1}{c}{\mgii\ \md } & \multicolumn{1}{c}{\mgii\ $v_{\rm min}$ } & \multicolumn{1}{c}{\mgii\ $v_{\rm max}$ } & \multicolumn{1}{c}{\mgii\ $v_{\rm cov,min}$ } & \multicolumn{1}{c}{\mgii\ $v_{\rm cov,max}$ } \\
    \multicolumn{1}{c}{ } & \multicolumn{1}{c}{ } & \multicolumn{1}{c}{ } & \multicolumn{1}{c}{ } & \multicolumn{1}{c}{ \kms } & \multicolumn{1}{c}{ } & \multicolumn{1}{c}{ \kms } & \multicolumn{1}{c}{ \kms } & \multicolumn{1}{c}{ \kms } & \multicolumn{1}{c}{ \kms } \\
    \hline
    0 & 0 & 0 & 0 &     0.0 & 0.000 &      0 &      0 &  -3000 & -25000 \\
    0 & 0 & 0 & 0 &     0.0 & 0.000 &      0 &      0 &  -3000 & -25000 \\
    1 & 0 & 0 & 0 &     0.0 & 0.000 &      0 &      0 &  -3000 & -25000 \\
    0 & 0 & 0 & 0 &     0.0 & 0.000 &      0 &      0 &  -3000 & -25000 \\
    0 & 0 & 0 & 0 &     0.0 & 0.000 &      0 &      0 &  -3000 & -25000 \\
    \hline
  \end{tabular}
\end{table*}

\begin{table*}
  \contcaption{}
  \begin{tabular}{rrrrrrrr}
    \hline
    \multicolumn{1}{c}{\mgii\ $N_{\rm BR}$ } & \multicolumn{1}{c}{\mgii\ $N_{\rm BT}$ } & \multicolumn{1}{c}{\mgii\ $N_{\rm SR}$ } & \multicolumn{1}{c}{\mgii\ $N_{\rm ST}$ } & \multicolumn{1}{c}{\mgii\ Inc } & \multicolumn{1}{c}{\mgii\ CBP } & \multicolumn{1}{c}{\mgii\ CPF } & \multicolumn{1}{c}{\mgii\ BBP } \\
    \multicolumn{1}{c}{ } & \multicolumn{1}{c}{ } & \multicolumn{1}{c}{ } & \multicolumn{1}{c}{ } & \multicolumn{1}{c}{ } & \multicolumn{1}{c}{ } & \multicolumn{1}{c}{ } & \multicolumn{1}{c}{ } \\
    \hline
      0 &   0 &  0 &  0 & 0 & 0 & 0 & 0 \\
      0 &   0 &  0 &  0 & 0 & 0 & 0 & 0 \\
      0 &   0 &  0 &  0 & 0 & 0 & 0 & 0 \\
      0 &   0 &  0 &  0 & 0 & 0 & 0 & 0 \\
      0 &   0 &  0 &  0 & 0 & 0 & 0 & 0 \\
    \hline
  \end{tabular}
\end{table*}

\end{document}